\documentclass{houches}

\usepackage[fleqn]{amsmath} 
\usepackage{epsfig}

\newcommand{\noi}{\noindent}

\newcommand{\ra}{\rightarrow}
\newcommand{\Ra}{\Rightarrow}

\newcommand{\lesssim}{ {\
\lower-1.2pt\vbox{\hbox{\rlap{$<$}\lower5pt\vbox{\hbox{$\sim$}}}}\ }  }
\newcommand{\gtrsim}{ {\
\lower-1.2pt\vbox{\hbox{\rlap{$>$}\lower5pt\vbox{\hbox{$\sim$}}}}\ }  } 

\newcommand{\cA}{{\cal A}}

\newcommand{\cC}{{\cal C}}

\newcommand{\cG}{{\cal G}}

\newcommand{\cL}{{\cal L}}
\newcommand{\cM}{{\cal M}}

\newcommand{\cO}{{\cal O}}

\newcommand{\cR}{{\cal R}}

\newcommand{\cU}{{\cal U}}

\newcommand{\CF}{C_{\rm F}}
\newcommand{\QCD}{\mbox{\rm {\tiny QCD}}}
\newcommand{\Imm}{\mbox{\rm Im}}
\newcommand{\Ree}{\mbox{\rm Re}}

\newcommand{\tr}{\mbox{\rm tr}}

\newcommand{\MeV}{\mbox{\rm MeV}}
\newcommand{\GeV}{\mbox{\rm GeV}}

\newcommand{\annd}{\mbox{\rm and}}

\newcommand{\alphak}{\alpha_{\mbox{\rm eff}}(k_{E}^2)}
\newcommand{\alpham}{\alpha(\mu^2)}
\newcommand{\alphaq}{\alpha(Q^2)}

\newcommand{\als}{\alpha_{\mbox{\rm {\scriptsize s}}}}
\newcommand{\gs}{g_{\mbox{\rm {\scriptsize s}}}}

\input epsf.sty

\begin{document}

\setcounter{chapter}{0}
\setcounter{page}{1}

\author[E.~de Rafael]{Eduardo de Rafael}
\address{Centre  de Physique Th\'eorique\\ CNRS-Luminy, Case 907\\
F-13288 Marseille Cedex 9, France\thanks{E-mail: EdeR@cpt.univ-mrs.fr}}

\chapter{An Introduction to Sum Rules in QCD }

\section{Introduction}

These lectures are an introduction to the subject of sum rules in hadronic
interactions as understood at present within the framework of quantum
chromodynamics (QCD). The literature on the subject is extensive and
various applications are quoted in other lectures at this school. I have
chosen to concentrate on the foundations of the subject rather than giving
an exhaustive list of results obtained from sum rules. The aim is to give
you a critical overview of the subject so that you can judge by yourselves
on how seriously to take the results in the literature, and also to foresee
which possible new applications could be made. One often encounters two
extreme attitudes among theorists on results obtained using QCD sum rules.
One is the optimistic belief that the result can be trusted at the few
percent level, when in fact there are a host of assumptions in the
derivation; the other is the pessimistic attitude which pretends that all
QCD sum rules are ``wrong'' because some calculation somebody made somewhere
gives an incorrect result. I want to show you that the existence of sum
rules in many cases follows rigorously from QCD. Often
the difficulty is not in the sum rule itself, but rather on how to use it in
practice, when one is dealing with a channel where there is very little
direct experimental information. We shall see that it is important to check
that the matching of the input low energy hadronic {\it ansatz} to the
continuum described by perturbative QCD (pQCD) satisfies {\it global
duality} tests.

The lectures are organized as follows. I shall start in
section~\ref{sec:srptqcd} with an introduction to sum rules which were
derived prior to the development of QCD. This section is mostly descriptive
and assumes some familiarity with topics like chiral perturbation theory
which are covered in other lectures~\cite{PiclesH,ManlesH} at this school,
and with dispersion relations. The reader who has never heard of dispersion
relations should first read the beginning of section~\ref{sec:disperrels} and
section~\ref{subsec:vpmf}, and then go back to section~\ref{sec:srptqcd}.
Section~\ref{sec:ttpfsr} contains quite a few technical details. It should
be useful to those who want to learn how to use QCD sum rules in practice.
Section~\ref{sec:nppc} gives a qualitative introduction to non--perturbative
power corrections. It should be helpful as an introduction to more technical
papers quoted there. The two examples discussed in section~\ref{sec:seqcdsr}
are rather topical subjects, much under discussion at the moment, and
illustrate how to use the material covered in the previous sections.

There are a lot of topics on sum rules which are not mentioned in these
introductory lectures like for example, applications to baryons, to heavy
quark systems and gluonium. The curious reader should complement these
lectures with extra reading. There are two references which contain a lot
of material. One is the book by Stephan Narison~\cite{Nar89}; the other is
M.A.~Shifman's selection of some of the original articles with his
comments in ref.~\cite{Shi92}.

\section{Sum Rules prior to QCD}\label{sec:srptqcd}

Sum rules in hadron physics have a long history. We shall be concerned in
this section with various sum rules which  were suggested before QCD was
established as the theory of the strong interactions. I want to do this for
two reasons: first, because they are interesting sum rules {\it per se} and
second, because they illustrate very clearly the basic ingredients which go
into the derivation of sum rules in general. The Drell--Hearn sum rule and
the Adler--Weisberger sum rule are expected to be true in QCD. As we shall
see later in section~\ref{subsec:wsrope}, the Weinberg sum rules are in fact
theorems of QCD in the chiral limit where the light quark masses are
neglected.  

\subsection{The Drell--Hearn Sum Rule}

The sum rule in question~\cite{DH66} relates the proton anomalous magnetic
moment $F_{2}(0)$ to an integral over the energy of the difference of total
photoabsorption polarized cross--sections  
\begin{equation}
\left[F_{2}(0)\right]^2 =
\frac{M_{{\mbox{\scriptsize\rm p}}}^2}{2\pi^2\alpha}\int_{\nu_{\mbox{\rm
th}}} ^{\infty}\frac{d\nu}{\nu}\left[\sigma_{\gamma{\mbox{\rm
p}}}^{\uparrow\uparrow} (\nu)-\sigma_{\gamma{\mbox{\rm
p}}}^{\uparrow\downarrow}(\nu)\right].
\end{equation} Here, $\sigma_{\gamma{\mbox{\scriptsize\rm
p}}}^{\uparrow\uparrow} (\nu) \left[
\sigma_{\gamma{\mbox{\scriptsize\rm p}}}^{\uparrow\downarrow}(\nu)\right]$
denote the total cross--section for the absorption of a circularly polarized
photon of laboratory energy $\nu$ by a proton polarized with its spin
parallel (antiparallel) to the photon spin,
$M_{{\mbox{\scriptsize\rm p}}}$ is the proton mass and $\alpha\simeq 1/137$
the fine structure constant. The integral starts at the inelastic threshold
$\nu_{{\mbox{\rm th}}}=m_{\pi}^2/2M_{\mbox{\scriptsize\rm p}}$. 

The three ingredients which are needed to derive this sum rule are: {\it
dispersion relations} for forward Compton--scattering, {\it the optical
theorem} which relates total cross--sections to the imaginary part of
forward scattering amplitudes, and a {\it low--energy theorem} which fixes
the value of the real forward scattering amplitude at zero energy. We shall
see these three ingredients appearing systematically in the derivation of
all the sum rules which we shall consider. However, in many of the QCD sum
rules which we shall later discuss the {\it low--energy theorem} will be
replaced by a {\it high--energy theorem} instead, or rather a {\it
short--distance behaviour} property which exploits the fact that two--point
functions in QCD are known perturbatively at high euclidean momentum
transfer due to the asymptotic freedom property of QCD.

\begin{figure}[htb]
\centerline{\epsfbox{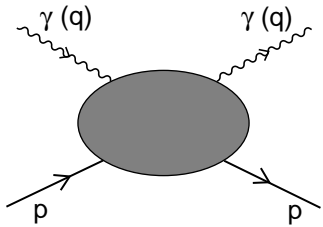}}
\caption{Forward Compton scattering of photons $\gamma(q)$ on protons $p$.}
\label{fig:Fig1a}
\end{figure}

The forward Compton--scattering amplitude illustrated by the diagram in
Fig.~\ref{fig:Fig1a},

\noi 
where $p\cdot q=M_{{\mbox{\scriptsize\rm p}}}\nu$, depends on two
scalar invariant amplitudes of the squared energy $\nu^2$:
\begin{equation} A(\nu)=f_{1}(\nu^2)\vec{\epsilon'}\cdot\vec{\epsilon}+
\nu f_{2}(\nu^2)i\vec{\sigma}\cdot\vec{\epsilon'}\times\vec{\epsilon}.
\end{equation} The spin--flip amplitude obeys a dispersion relation which is
expected to have no subtractions i.e.,
\begin{equation}
\Ree f_{2}(\nu^2)={\mbox{\rm
PP}}\!\int_{0}^{\infty}\frac{d\tilde{\nu}^{2}}{\tilde{\nu}^{2}-\nu^2}
\,\frac{1}{\pi}\Imm
f_{2}(\tilde{\nu}^{2}).
\end{equation} 
[Section~\ref{sec:disperrels} will be dedicated to the study of
dispersion relations in general. They are the basics of sum rule
derivations.] The {\it optical theorem} relates $\Imm f_{2}(\nu^2)$ to
total cross--sections:
\begin{equation} 
8\pi\,\Imm f_{2}(\nu^2)=\sigma_{\gamma{\mbox{\rm
p}}}^{\uparrow\downarrow} (\nu)-\sigma_{\gamma{\mbox{\rm
p}}}^{\uparrow\uparrow}(\nu),
\end{equation} 
and the {\it low energy--theorem} relates $\Ree f_{2}(0)$ to
the proton anomalous magnetic moment squared:
\begin{equation}
f_{2}(0)=-\frac{1}{2}\frac{\alpha}{M_{{\mbox{\scriptsize\rm
p}}}^2}\left[F_{2}(0)\right]^2.
\end{equation}

The {\it optical theorem} is a general property of unitarity in quantum field
theory. The {\it low--energy theorem} follows from chiral symmetry properties
of the underlying QCD Lagrangian~[{\footnotesize See the lectures of
Pich~\cite{PiclesH} and Manohar~\cite{ManlesH} in this school.}] The
{\it dispersion relation} for forward scattering amplitudes is also a general
property of quantum field theory~\cite{GGT54}. What is still lacking to
promote this sum rule to the status of a QCD theorem is the proof that
$\Imm f_{2}(\nu^2)\ra 0$ when $\nu\ra\infty$. At present this accepted
behaviour is only supported by theoretical strong interaction wisdom, like
Regge phenomenology. 

The Drell--Hearn sum rule has been
checked in perturbation theory in QED--like theories to the first
non--trivial order. One may ask if this sum rule could be used for practical
calculations of the anomalous magnetic moment of the electron. The answer
is that it is not a competitive method, simply because
$F_{2}(0)$ in the l.h.s. appears squared which obliges one to calculate the
integral in the r.h.s. to very high orders in powers of the fine structure
constant. 

The Drell--Hearn sum rule is also useful in the phenomenological study of
deep inelastic scattering of polarized leptons on polarized
protons~[{\footnotesize See Manohar's lectures~\cite{ManlesH} in this
school.}], where it provides a constraint on the limit of real
photoabsorption. It plays an important r\^{o}le as well in the calculation
of the so called {\it polarizability} contribution to the hyperfine
splitting in the hydrogen atom~[{\footnotesize See e.g. ref.~\cite{LPR72}
and references therein.}]

\subsection{The Adler--Weisberger Sum Rule}

\begin{figure}[htb]
\centerline{\epsfbox{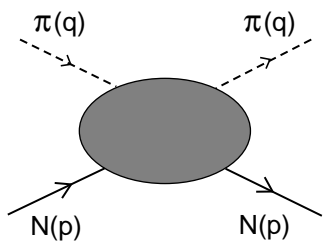}}
\caption{Forward pion--nucleon scattering amplitude.}
\label{fig:Fig1b}
\end{figure}

The sum rule in question~\cite{Adl65,Wei66} relates the axial coupling
constant $g_{\mbox{\scriptsize\rm A}}$ of $\beta$--decay to integrals of
pion--nucleon total cross--sections: 
\begin{equation} 1-g_{\mbox{\scriptsize\rm
A}}^2=\frac{2f_{\pi}^2}{\pi}\int_{0}^{\infty}\frac{d\nu}{\nu}
\left[\sigma_{\mbox{\scriptsize\rm tot}}^{\pi^{-}{\mbox{\scriptsize\rm
p}}}(\nu)-
\sigma_{\mbox{\scriptsize\rm tot}}^{\pi^{+}{\mbox{\scriptsize\rm p}}}(\nu)
\right],
\end{equation} where $f_{\pi}\simeq 92\,\MeV$ denotes the pion decay
coupling constant. The reference amplitude in this case is the forward
pion--nucleon scattering amplitude illustrated in Fig.~\ref{fig:Fig1b}

\begin{figure}[htb]
\centerline{\epsfbox{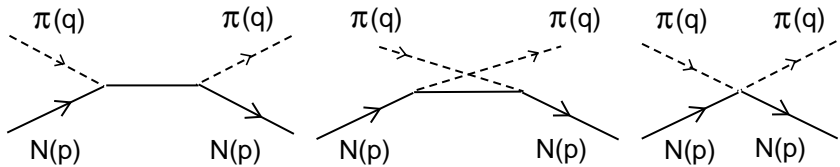}}
\caption{Forward pion--nucleon scattering at the Born
approximation.}
\label{fig:Fig1c}
\end{figure}

\noi
which has two invariant amplitudes:
\begin{equation}
T_{ij}(\nu)=\delta_{ij}\,T^{(+)}(\nu)+\frac{1}{2}\left[\tau_{i},
\tau_{j}\right]
\,\frac{M_{\mbox{\scriptsize\rm p}}\nu}{f_{\pi}^2}T^{(-)}(\nu),
\end{equation}
where $i,j$ are isospin indices. 
The basic dispersion relation is the one obeyed by the isospin--odd
amplitude
$T^{(-)}(\nu)$, and as in the case of the Compton scattering spin--flip
amplitude, it is expected that $T^{(-)}(\nu)$ obeys an unsubtracted
dispersion relation. The optical theorem relates
$\Imm T^{(-)}(\nu)$ to the difference of $\pi^-$--p and
$\pi^+$--p total cross--sections. On the other hand, the calculation of the
low--energy behaviour of the amplitude $\Ree T^{(-)}(\nu)$ can be done
using baryon chiral perturbation theory~\cite{ManlesH,PiclesH}. The
relevant Feynman diagrams are those of the Born approximation shown in
Fig.~\ref{fig:Fig1c}

\noi 
with the result
\begin{equation}
\lim_{\nu\ra 0}\Ree T^{(-)}(\nu)=1-g_{\mbox{\scriptsize\rm A}}^2.
\end{equation}
The seagull graph contributes the factor 1, each one of the other graphs a
factor $-\frac{1}{2}g_{\mbox{\scriptsize\rm A}}^2$.

Sometimes, the Adler--Weisberger sum rule is also written in the following
way
\begin{equation} 1-\frac{1}{g_{\mbox{\scriptsize\rm
A}}^2}=\frac{2M_{\mbox{\scriptsize\rm N}}^2} {\pi g_{\pi\mbox{\scriptsize\rm
N}\mbox{\scriptsize\rm N}}^2}\,
\int_{0}^{\infty}\frac{d\nu}{\nu}
\left[\sigma_{\mbox{\scriptsize\rm tot}}^{\pi^{+}
{\mbox{\scriptsize\rm p}}}(\nu)-
\sigma_{\mbox{\scriptsize\rm tot}}^{\pi^{-}{\mbox{\scriptsize\rm p}}}(\nu)
\right].
\end{equation} 
Both forms are equivalent, because of the Goldberger--Treiman
relation~\cite{GT58}
\begin{equation} g_{\mbox{\scriptsize\rm A}}M_{\mbox{\scriptsize\rm N}}=
g_{\pi\mbox{\scriptsize\rm N}\mbox{\scriptsize\rm N}}f_{\pi}.
\end{equation} The Goldberger--Treiman relation is also a low--energy
theorem of QCD.

I would like to come back to the question of an unsubtracted dispersion
relation for the isospin--odd amplitude $T^{(-)}(\nu)$. This, of course
depends on the high--energy behaviour of the underlying theory. One can
consider as a toy model  the linear sigma model~\cite{GL60} for constituent
quarks. It has been shown~\cite{Per91,Per92} that in perturbation theory,
this model does not require subtractions in the corresponding
Adler--Weisberger sum rule and yields, at the one--loop level, the result
\begin{equation}\label{eq:lsm}
g_{\mbox{\scriptsize\rm A}}^2=1-2\frac{M_{Q}^2}{16\pi^2 f_{\pi}^2}\left(
\log\frac{M_{\sigma}^2}{M_{Q}^2}+\cO (1)\right),
\end{equation}
where $M_{Q}$ denotes the mass of the constituent quark in this model. On
the other hand, if one considers an effective constituent chiral quark model
like the Georgi--Manohar model~\cite{GM84}, then it can be
shown~\cite{PdeR93} that the corresponding isospin--odd amplitude $T^{(-)}$
does not obey an unsubtracted dispersion relation. Beyond the tree level,
$\Ree T^{(-)}(\nu)$ cannot be obtained from a dispersive integral; and
indeed, the calculation at the one--loop level reproduces the result in
eq.~(\ref{eq:lsm}) but with a cut--off replacing the sigma mass. It is
simply that in this case the effective low--energy theory does not have
the high--energy behaviour of the underlying theory.

\subsection{The Weinberg Sum Rules}\label{subsec:wsrss}

\begin{figure}[htb]
\centerline{\epsfbox{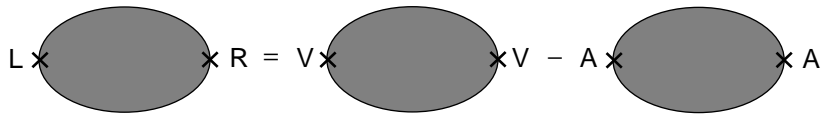}}
\caption{The $LR$ two--point function is the difference of $VV$ and
$AA$ two--point functions.}
\label{fig:Fig1d}
\end{figure}

There is a correlation function  which is particularly sensitive to
properties of chiral symmetry breaking, namely the
two--point function 
\begin{equation}\label{eq:lrtpf}
\Pi_{LR}^{\mu\nu}(q)=2i\int d^4 x\,e^{iq\cdot x}\langle 0\vert
\mbox{\rm T}\left(L^{\mu}(x)R^{\nu}(0)^{\dagger}\right)\vert 0\rangle ,
\end{equation} 
with currents
\begin{equation} 
L^{\mu}=\bar{d}(x)\gamma^{\mu}\frac{1}{2}(1-\gamma_{5})u(x)
\quad \annd \quad 
R^{\mu}=\bar{d}(x)\gamma^{\mu}\frac{1}{2}(1+\gamma_{5})u(x)\,.
\end{equation} 
In the chiral limit where the light quark masses are set to
zero, this two--point function only depends on one invariant amplitude
($Q^2\equiv -q^2\ge 0$ for $q^2$ spacelike)
\begin{equation}\label{eq:lritpf}
\Pi_{LR}^{\mu\nu}(Q^2)=(q^{\mu}q^{\nu}-g^{\mu\nu}q^2)\Pi_{LR}(Q^2).
\end{equation} 
The function $\Pi_{LR}(Q^2)$ vanishes order by order in
perturbation theory and is an order parameter of the spontaneous breakdown
of chiral symmetry (S$\chi$SB) for all values of the momentum transfer. As
illustrated in Fig.~\ref{fig:Fig1d}

\noi the two--point function in eq.~(\ref{eq:lrtpf}) is the difference of a
vector current correlation function and an axial--vector current correlation
function. Chiral symmetry identifies these two correlation functions. 

Two--point functions in general obey  dispersion relations.~[{\footnotesize
See the discussion in section~\ref{sec:disperrels}.}] In 1967, and therefore
prior to the development of QCD, Weinberg made an educated guess
about the convergence properties of the LR two--point function, with the
suggestion that the following two sum rules may hold in the underlying
theory of the strong interactions~\cite{We67}:
\begin{equation}\label{eq:1stwsr}
\int_{0}^{\infty}dt\Imm\Pi_{LR}(t)\equiv
\int_{0}^{\infty}dt\left[\Imm\Pi_{V}(t)-\Imm\Pi_{A}(t)\right]=0,
\end{equation} and
\begin{equation}\label{eq:2ndwsr}
\int_{0}^{\infty}dt\,t\Imm\Pi_{LR}(t)\equiv
\int_{0}^{\infty}dt\,t\left[\Imm\Pi_{V}(t)-\Imm\Pi_{A}(t)\right]=0.
\end{equation} They are commonly referred to as the {\it 1st Weinberg sum
rule} and the {\it 2nd Weinberg sum rule}, and they are examples of the so
called {\it superconvergent} sum rules for the following reason. Let us
assume that the function
$\Pi_{LR}(Q^2)$ obeys an unsubtracted dispersion relation:
\begin{equation}\label{eq:udrwsr}
\Pi_{LR}(Q^2)=\int_{0}^{\infty}\frac{dt}{t+Q^2}\Imm\Pi_{LR}(t).
\end{equation} 
The existence of the two sum rules above implies a stronger
convergence behaviour for the function $\Pi_{LR}(Q^2)$ in the deep
euclidean region than the minimum required for the unsubtracted dispersion
relation to hold. Expanding
$\frac{1}{t+Q^2}=\frac{1}{Q^2}-
\frac{1}{Q^2}t\frac{1}{Q^2}+\cdots$, we observe that the {\it 1st Weinberg
sum rule} corresponds to the constraint
\begin{equation}\label{eq:ope2}
\lim_{Q^2\ra\infty} Q^2\Pi_{LR}(Q^2)\Ra 0,
\end{equation} 
and the {\it 2nd Weinberg sum rule} to the further constraint
\begin{equation}\label{eq:ope4}
\lim_{Q^2\ra\infty} Q^4\Pi_{LR}(Q^2)\Ra 0.
\end{equation} 
It is quite remarkable that, as we shall see in
section~\ref{subsec:wsrope}, both constraints are now QCD properties; and in
fact, the {\it 1st Weinberg sum rule} still holds in the presence of finite
light quark masses~\cite{FNR79}.

I would like to make some comments on these sum rules. 
 
i) Weinberg in his original paper~\cite{We67} assumed that only the
lowest narrow resonant state in the vector and axial--vector spectral
functions:
\begin{equation}\label{eq:Vresdom}
\frac{1}{\pi}\Imm\Pi_{V}(t)=f_{V}^2 M_{V}^2\delta(t-M_{V}^2)+\cdots
\end{equation} and
\begin{equation}\label{eq:Aresdom}
\frac{1}{\pi}\Imm\Pi_{A}(t)=f_{\pi}^2\delta(t)+ f_{A}^2
M_{A}^2\delta(t-M_{A}^2)+\cdots ,
\end{equation} contribute significantly to the sum rules. The first term in
the axial--vector spectral function in eq.~(\ref{eq:Aresdom}) is the
contribution from the massless pion. Thus the {\it 1st and 2nd Weinberg sum
rules} in eqs.(\ref{eq:1stwsr}) and (\ref{eq:2ndwsr})
constrain  the couplings and masses of the narrow resonances as
follows
\begin{equation}\label{eq:tsc}
\begin{split} f_{V}^2 M_{V}^2 -f_{A}^2 M_{A}^2 &= f_{\pi}^2 \\ f_{V}^2
M_{V}^4 -f_{A}^2 M_{A}^4 &= 0.
\end{split}
\end{equation}
It is interesting that the original Weinberg's {\it ansatz} in
eqs.~(\ref{eq:Vresdom}) and (\ref{eq:Aresdom})  can now be justified  within
the framework of the large--$N_c$ limit of QCD. Indeed, in this limit, the
vector and axial--vector spectral functions are each a sum of an infinite
number of narrow sates. The dots in the spectral functions in
eqs.~(\ref{eq:Vresdom}) and (\ref{eq:Aresdom}) implicitly assume that the
sum of the rest of the narrow states is already globally dual to the onset
of the perturbative continuum, 
\begin{equation}\label{eq:Vresdomqcd}
\frac{1}{\pi}\Imm\Pi_{V}(t)=f_{V}^2 M_{V}^2\delta(t-M_{V}^2)+
\frac{N_c}{16\pi^2}\frac{2}{3}\theta(t-s_{0})[1+\cdots],
\end{equation} 
and
\begin{equation}\label{eq:Aresdomqcd}
\frac{1}{\pi}\Imm\Pi_{A}(t)=f_{\pi}^2\delta(t)+ f_{A}^2
M_{A}^2\delta(t-M_{A}^2)+\frac{N_c}{16\pi^2}\frac{2}{3}
\theta(t-s_{0})[1+\cdots],
\end{equation} 
where the dots indicate now gluonic corrections. The
vector and axial--vector spectral functions of the pQCD continuum are the
same in the chiral limit.

ii) In order to make a ``prediction'' for the axial mass $M_{A}$,
Weinberg in his original paper also makes the assumption that
\begin{equation} f_{V}^2=2\frac{f_{\pi}^2}{M_{V}^2},
\end{equation} which was called at the time the KSFR
relation~\cite{KS66,FR66}. This, together with eqs.~(\ref{eq:tsc}) leads to
the prediction
$M_{A}=\sqrt{2}M_{V}$ which is satisfied by the central values of the
presently known masses within an accuracy of less than $\sim 12\%$. It can
be shown~\cite{EGLPR89} that the KSFR--relation follows from the
assumption that both the electromagnetic pion form factor and the axial form
factor in
$\pi\ra e\nu_{e}\gamma$ obey unsubtracted dispersion relations, plus the
dynamical input of resonance dominance.

iii) Inverse moments of the difference of vector and axial--vector
spectral functions with the pion pole removed, i.e. integrals like
\begin{equation}
\int_{0}^{\infty}dt\frac{1}{t^p}\left(\frac{1}{\pi}\Imm\Pi_{LR}(t)
+f_{\pi}^2\delta(t)\right),
\end{equation} 
with $p=1,2,3,\cdots$, correspond to non--local order
parameters which govern the couplings of local operators of higher and
higher dimensions in the low energy effective chiral
Lagrangian~[{\footnotesize See Pich's lectures~\cite{PiclesH} in this
school.}]. For example, the first inverse moment is related to the
$\cO(p^4)$ coupling constant $L_{10}$ as follows~\cite{GL84}
\begin{equation} 
-4L_{10}=\int_{0}^{\infty}\frac{dt}{t}\,
\left(\frac{1}{\pi}\Imm\Pi_{V}(t)-
\frac{1}{\pi}
\Imm\tilde{\Pi}_{A}(t)\right),
\end{equation}
where the tilde in $\Imm\tilde{\Pi}$ indicates that the pion pole has
been removed. Using the ansatz in eqs.~(\ref{eq:Vresdomqcd}) and
(\ref{eq:Aresdomqcd}) and the two Weinberg sum rules in eq.~(\ref{eq:tsc})
one finds
\begin{equation}
-4L_{10}=f_{V}^2-f_{A}^2=f_{\pi}^2\left(\frac{1}{M_{V}^2}+\frac{1}
{M_{A}^2}\right),
\end{equation} which provides a good estimate~\cite{EGLPR89} of the
$L_{10}$ coupling constant.

iv)  One of the parameters which characterize possible deviations from
the Standard Model predictions in the electroweak sector is the so called
$S$--parameter~[{\footnotesize See the lectures of Treille~\cite{TrelesH}
and Chivukula~\cite{ChilesH} in this school.}]. In the unitary gauge,
$S$ measures the strength of an anomalous $W^{(3)}_{\mu\nu}B^{\mu\nu}$
coupling. This is the
$SU(2)_{L}\times SU(2)_{R}$ coupling analogous to the term proportional to
$L_{10}$ in QCD. It has been argued that if the underlying theory of
electroweak breaking is a vector--like gauge theory of the QCD--type, the
sign of the
$L_{10}$--like coupling must be negative, precisely as observed in QCD. In
the electroweak sector, this fact is infirmed  by experimental observation
and constitutes at present a serious phenomenological obstacle to
technicolour--like formulations of electroweak symmetry breaking. The
relation of the
$L_{10}$ coupling to the relative size of local order parameters versus
$f_{\pi}^2$ in the large--$N_c$ limit of QCD--like theories has been
recently discussed in ref.~\cite{KdeR97}.

\subsection{The Electromagnetic Pion Mass Difference}

The most remarkable application of the Weinberg sum rules is the
calculation of the $\pi^{+}-\pi^{0}$ electromagnetic mass
difference~\cite{Lowetal67}. In the chiral limit and to lowest order in the
electromagnetic interactions, there appears a new term with no derivatives
in the low--energy effective chiral Lagrangian~\cite{EGPR89}
\begin{equation}\label{eq:effem}
\cL_{\mbox{\scriptsize\rm eff}}=\cdots +e^2 C\,\tr
\left(Q_{R}UQ_{L}U^{\dagger}\right),
\end{equation}
where $U$ is the matrix field which collects the octet of pseudoscalar
Goldstone fields~[{\footnotesize See Pich's lectures~\cite{PiclesH} in this
school.}] and $Q_{R}=Q_{L}=\mbox{\rm diag}[2/3, -1/3, -1/3]$ the right--
and left--charges associated with the electromagnetic couplings of the light
quarks. Expanding $U$ in powers of the pseudoscalar fields there appear
quadratic terms like
\begin{equation}\label{eq:effemex}
\cL_{\mbox{\scriptsize\rm eff}}=\cdots -2e^2 C\frac{1}{f_{\pi}^2} (\pi^+
\pi^- + K^+ K^-) + \cdots ,
\end{equation}  
showing that, in the presence of electromagnetic
interactions, the charged pion and kaon fields become massive:
\begin{equation} 
(m_{K^+}^{2}-m_{K^0}^{2})\vert_{\mbox{\scriptsize\rm
EM}}=(m_{\pi^+}^{2}-m_{\pi^0}^{2})\vert_{\mbox{\scriptsize\rm
EM}}=
\frac{2e^2 C}{f_{\pi}^2}.
\end{equation}
In fact, the main contribution to the physical $\pi^{+}$--$\pi^{0}$ mass
difference is of electromagnetic origin because quark masses do not
contribute significantly to the $\pi^{+}$--$\pi^{0}$ mass splitting.

It can be
formally  shown~[{\footnotesize See e.g. refs.~\cite{BdeR91,KU97} and
references therein.}] that the coupling constant
$C$ of the effective term in eq.~(\ref{eq:effem}) 
which results from the integration of a virtual photon of euclidean
momentum squared $Q^2$ in the presence of the strong interactions in the
chiral limit, is given by an integral of the correlation function
$\Pi_{LR}(Q^2)$ defined in eqs.~(\ref{eq:lrtpf}) and (\ref{eq:lritpf}):
\begin{equation}\label{eq:piem} 
C=
\frac{-1}{8\pi^2}\,\frac{3}{4}\,
\int_0^\infty dQ^2\,Q^2\Pi_{LR}(Q^2).
\end{equation}
The constraints in eqs.~(\ref{eq:ope2}) and (\ref{eq:ope4}) guarantee the
convergence of the integral in the ultraviolet.
Inserting the resonance dominance ansatz in
eqs.~(\ref{eq:Vresdomqcd}) and (\ref{eq:Aresdomqcd}) for the vector and
axial--vector spectral functions in the dispersion integral representation
in eq.~(\ref{eq:udrwsr}) results in 
\begin{equation}\label{eq:lrtpfrd}
\Pi_{LR}(Q^2)=-\frac{f_{\pi}^2}{Q^2}-\frac{f_{A}^2 M_{A}^2} {M_{A}^2
+Q^2}Ê+ \frac{f_{V}^2 M_{V}^2} {M_{V}^2 +Q^2}.
\end{equation} 
With this expression inserted in the integrand of the r.h.s. in
eq.~(\ref{eq:piem}) and using the 1st and 2nd Weinberg sum rules in
eqs.~(\ref{eq:tsc}) one easily obtains
\begin{equation} 
(m_{\pi^+}^{2}-m_{\pi^0}^{2})\vert_{\mbox{\scriptsize\rm
EM}}=
\frac{\alpha}{\pi}\frac{3}{4}\frac{M_{A}^2 M_{V}^2}{M_{A}^2 -M_{V}^2}
\log\frac{M_{A}^2}{M_{V}^2}.
\end{equation} 
The authors of ref.~\cite{Lowetal67} also use
$M_{A}=\sqrt{2}M_{V}$ which results in
\begin{equation} (m_{\pi^+}^{2}-m_{\pi^0}^{2})\vert_{\mbox{\scriptsize\rm
EM}}=
\frac{\alpha}{\pi}\frac{3}{2}\,M_{\rho}^2\,\log 2.
\end{equation} Numerically, this corresponds to a mass difference
$\Delta m_{\pi}\vert_{\mbox{\scriptsize\rm EM}}\simeq 5.2\,\MeV$ to be
compared with the experimental mass difference
$m_{\pi^+}-m_{\pi^0}=4.6\,\MeV$.

Equation (\ref{eq:piem}) can nowadays be viewed as a rigorous QCD sum rule
in the chiral limit. The sum rule relates the low--energy constant
$(m_{\pi^+}^{2}-m_{\pi^0}^{2})\vert_{\mbox{\scriptsize\rm EM}}$ to an
integral of the real part of a two--point function. The integral runs over
all possible values of the euclidean momenta and therefore it requires, a
priori, the knowledge of the two--point function at all distances. I want
to stress the fact that it is not enough to know the very low $Q^2$
behaviour of the $\Pi_{LR}(Q^2)$ correlation function to obtain an
approximation to the r.h.s. integral in eq~(\ref{eq:piem}). The low $Q^2$
behaviour of this function is governed by chiral perturbation theory and
the first two terms are known
\begin{equation}
\Pi_{LR}(Q^2)=-\frac{f_{\pi}^2}{Q^2}-4L_{10}+\cO(Q^2);
\end{equation} but this behaviour cannot be extrapolated to the ultraviolet
region of the integral since it would lead to a divergent result. By
contrast, as we shall further discuss in section~\ref{subsec:wsrope}, the
full
$Q^2$ resonance dominance ansatz in eq.~(\ref{eq:lrtpfrd}) with the Weinberg
sum rule constraints in eqs.~(\ref{eq:tsc}) has both the long--distance
behaviour and the short--distance behaviour of QCD.

\section{Dispersion Relations}\label{sec:disperrels}

In these lectures we shall be all the time considering  two--point
functions of local operators i.e., Fourier transforms of the vacuum
expectation value of the time ordered product of a local current
$J(x)$ times its hermitian conjugate
\begin{equation}\label{eq:tpf}
\Pi(q^2)=i\int d^{4}x e^{iq\cdot x}\langle 0\vert
\mbox{\rm T}\left\{J(x)J(0)^{\dagger}\right\}\vert 0\rangle.
\end{equation} 
In most applications the current $J(x)$ is one of the Noether
currents associated with global gauge transformations of flavour degrees of
freedom, like a vector current
$\bar{q}(x)\gamma^{\mu}q(x)$, or an axial--vector current
$\bar{q}(x)\gamma^{\mu}\gamma_{5}q(x)$; but one could also consider gauge
invariant local operators of gluon fields like $\tr
G^{\mu\nu}(x)G_{\mu\nu}(x)$, or composite operators like
$\bar{q}(x)\Gamma q(x)\bar{q}(x)\Gamma q(x)$. It has been shown by
K\"{a}ll\'{e}n and Lehmann~\cite{Ka52,Le54} quite a long time ago that
two--point functions obey {\it dispersion relations}. The dispersion
relation follows from the analyticity properties of $\Pi(q^2)$ as a complex
function of $q^2$, the only energy--momentum invariant which appears in a
two--point function. In full generality
$\Pi(q^2)$ is an analytic function in the complex $q^2$--plane but for a
cut in the real axis $0\le q^2 \le\infty$. It then follows that
\begin{equation}\label{eq:selfdisp}
\Pi(q^2)=\int_{0}^{\infty}dt\frac{1}{t-q^2
-i\epsilon}\frac{1}{\pi}\Imm\Pi(t) + \mbox{\rm a} + \mbox{\rm b}q^2 +
\cdots,
\end{equation} 
where  the degree of the arbitrary polynomial in the r.h.s.
depends on the convergence properties of $\Imm\Pi(t)$ when
$t\ra\infty$. The interest of this representation is that
$\frac{1}{\pi}\Imm\Pi(t)$ in the integrand, which is usually called the
{\it spectral function}, is a physical cross--section. With
$J(x)$ a current with specific quantum numbers, the spectral function is
then directly related to the total cross section for the production from
the vacuum of hadronic states with those quantum numbers.

For example, with $J$ the electromagnetic hadronic current of light quarks,
\begin{equation} 
J^{\mu}(x)=\frac{2}{3}\bar{u}(x)\gamma^{\mu}u(x)-
\frac{1}{3}\bar{d}(x)\gamma^{\mu}d(x)-
\frac{1}{3}\bar{s}(x)\gamma^{\mu}s(x)
\end{equation} 
the relation to the total $e^+ e^-$ annihilation
cross--section into hadrons is
\begin{equation}
\sigma(q^2)_{e^+ e^- \ra\mbox{\rm hadrons}}=\frac{4\pi^2 \alpha}{q^2}
\,e^2 \frac{1}{\pi}\Imm\Pi_{\mbox{\tiny\rm EM}}(q^2),
\end{equation} 
and
\begin{equation}\label{eq:spechad}
\begin{split} 
-3\theta(q)q^2 &\,\frac{1}{\pi}\Imm\Pi_{\mbox{\tiny\rm
EM}}(q^2)= \\
&\sum_{\Gamma}\langle 0\vert J^{\mu}_{\mbox{\tiny\rm EM}}(0)\vert
\Gamma\rangle
\langle\Gamma\vert J_{\mu,\mbox{\tiny\rm EM}}(0)^{\dagger}\vert 0\rangle
(2\pi)^{3}\delta^{(4)}(q-p_{\Gamma}),
\end{split}
\end{equation} 
where the sum is extended to all possible physical states,
{\it on--shell states}, with an integration over their corresponding phase
space understood. A pictorial representation of these equations is shown in
Fig.~\ref{fig:Fig2a} below.
 
\begin{figure}[htb]
\centerline{\epsfbox{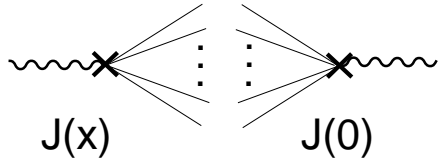}}
\caption{Representation of the hadronic spectral function in
eq.~(\ref{eq:spechad}).}
\label{fig:Fig2a}
\end{figure}
\begin{figure}[htb]
\centerline{\epsfbox{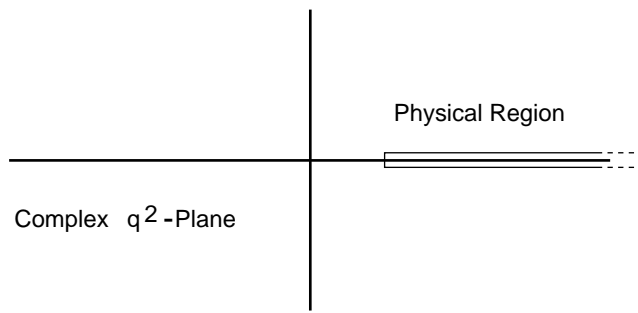}}
\caption{Representation of the complex $q^2$--plane.}
\label{fig:Fig2b}
\end{figure}

\noi
The lowest possible state in this case is a two--pion state.
The function $\Pi_{\mbox{\tiny\rm EM}}(q^2)$ is analytic in the complex
$q^2$--plane but for a cut in the real axis:
$4m_{\pi}^2\le q^2\le\infty$, as illustrated in Fig.~\ref{fig:Fig2b}.

I want to show first how these properties appear explicitly in a simple
case which you should be familiar with,  and then we shall discuss a
general proof.  

\subsection{Vacuum Polarization due to Massive Fermions}\label{subsec:vpmf}

\begin{figure}[htb]
\centerline{\epsfbox{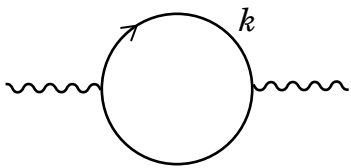}}
\caption{One fermion loop contribution to vacuum polarization.}
\label{fig:Fig2c}
\end{figure}

Let us consider QED with a massive fermion. We want to calculate the lowest
order vacuum polarization contribution from this fermion, i.e. the
one--loop Feynman diagram in Fig.~\ref{fig:Fig2c}.

\noi Combining the two fermion propagators
with a Feynman parameter $x$,
\begin{equation}
\frac{1}{AB}=\int_{0}^{1}\frac{1}{[Ax+B(1-x)]^2},
\end{equation} 
and after integration over the four--momenta $k$ in the
loop [{\footnotesize I am assuming that a gauge invariant
regularization of this integral, like e.g. dimensional regularization, has
been made.}], the one--particle irreducible part of this diagram, when
subtracted at
$q^2=0$ if we choose on--shell renormalization, leads to the renormalized
self--energy function:
\begin{equation}\label{eq:selffeynman}
\Pi_{\mbox{\tiny\rm R}}(q^2)=-\frac{\alpha}{\pi}\int_{0}^{1}
dx\,2x(1-x)\log\left(1-\frac{q^2}{m^2-i\epsilon}x(1-x)\right),
\end{equation} 
With the change of variables $y=1-2x$, and using the fact
that the resulting integral is symmetric when $y\ra -y$, we get
\begin{equation}
\Pi_{\mbox{\tiny\rm R}}(q^2)=\frac{\alpha}{\pi}\int_{0}^{1}
dy(1-y^2)\log\left[1-\frac{q^2}{4m^2 -i\epsilon}(1-y^2)\right].
\end{equation} 
Integrating by parts this equation using the identity:
$1-y^2=\frac{\partial}{\partial y}(y-\frac{1}{3}y^3)$ results in a rational
integral:
\begin{equation}
\Pi_{\mbox{\tiny\rm R}}(q^2)=\frac{\alpha}{\pi}\int_{0}^{1}dy\,
2y(y-\frac{1}{3}y^3)\frac{q^2}{4m^2-q^2(1-y^2)-i\epsilon},
\end{equation} 
and with a new change of variables: $t=\frac{4m^2}{1-y^2}$ we
finally get a rather interesting representation of the renormalized
self--energy
\begin{equation}\label{eq:fdisprel}
\frac{\Pi_{\mbox{\tiny\rm
R}}(q^2)}{q^2}=\frac{\alpha}{\pi}\int_{4m^2}^{\infty}
\frac{dt}{t}\,\frac{1}{t-q^2-i\epsilon}\,
\frac{1}{3}\left(1+\frac{2m^2}{t}\right)\sqrt{1-\frac{4m^2}{t}}.
\end{equation} 
The reason why this representation is interesting is that it
is in fact a dispersion relation! We have succeeded in rewriting the initial
Feynman parametric representation in eq.~(\ref{eq:selffeynman}) as a
dispersion relation by simple changes of variables. Using the identity
\begin{equation}\label{eq:ppartid}
\frac{1}{t-q^2-i\epsilon}=\mbox{\rm PP}\frac{1}{t-q^2} +i\pi\delta(t-q^2),
\end{equation} 
we immediately see that
\begin{equation}
\frac{1}{\pi}\Imm\Pi(t)=\frac{\alpha}{\pi}
\frac{1}{3}\left(1+\frac{2m^2}{t}\right)\sqrt{1-\frac{4m^2}{t}}
\theta(t-4m^2).
\end{equation} 
Equation (\ref{eq:fdisprel}) is a particular case of the
general dispersion relation written in eq.~(\ref{eq:selfdisp}), when the
arbitrary polynomial is just a constant, and we have got rid of the
constant because the on--shell renormalized
$\Pi_{\mbox{\tiny\rm R}}$ is defined as 
\begin{equation}
\Pi_{\mbox{\tiny\rm R}}(q^2)=\Pi(q^2)-\Pi(0)=
\int_{0}^{\infty}
\frac{dt}{t}\,\frac{q^2}{t-q^2-i\epsilon}\,\frac{1}{\pi}\Imm\Pi(t).
\end{equation}

It is perhaps worth insisting on the fact that asymptotically
\begin{equation}
\lim_{t\ra\infty}\frac{1}{\pi}\Imm\Pi(t)\Ra\frac{\alpha}{\pi}
\frac{1}{3};
\end{equation} 
i.e. the electromagnetic spectral function goes to a
constant. In fact, it is this constant which fixes the value of the lowest
order contribution to the
$\beta$--function associated with charge renormalization in QED.

\subsection{General Proof of Dispersion Relations}

We shall now sketch a proof of the dispersion relation property for
two--point functions in full generality. The key of the proof lies in the
definition of the time ordered product implicit in eq.~(\ref{eq:tpf}):
\begin{equation}
\mbox{\rm T}(J(x)J(0)^{\dagger}=\theta(x)J(x)J(0)^{\dagger}+
\theta(-x)J(0)^{\dagger}J(x),
\end{equation} 
and the use of translation invariance. The function
$\theta(x)$ is the Heaviside function: $\theta(x)=1$ if $x_{0}>0$ and
$\theta(x)=0$ if
$x_{0}<0$, which has the integral representation
\begin{equation}\label{eq:theta}
\theta(x)=\frac{1}{2\pi i}\int_{-\infty}^{+\infty}dw\,
\frac{ e^{iwx}}{w-i\epsilon}.
\end{equation}

First we insert a complete set of states
$\sum_{\Gamma}\vert\Gamma\rangle
\langle\Gamma\vert$ between the two currents in the T--product definition.
This leads to matrix elements of the type $\langle 0\vert
J(x)\vert\Gamma\rangle$ to which we apply translation invariance: 
\begin{equation}
\langle 0\vert J(x)\vert \Gamma\rangle=\langle 0\vert\cU^{-1}\cU
J(x)\cU^{-1}\cU\vert\Gamma\rangle,
\end{equation} 
where $\cU$ is the unitary operator induced by translations
in space--time:
\begin{equation}
\cU(a)J(x)\cU^{-1}(a)=J(x+a)\quad\annd\quad\cU(a)\vert\Gamma\rangle=
e^{ip_{\Gamma}\cdot a}\vert\Gamma\rangle,
\end{equation} 
and $p_{\Gamma}$ denotes the sum of the energy--momenta
of all the particles which define the state $\vert\Gamma\rangle$. The choice
$a\equiv -x$ factors out the
$x$--dependence of the matrix element into an exponential:
\begin{equation}
\langle 0\vert J(x)\vert\Gamma\rangle=e^{-ip_{\Gamma}\cdot x}\,
\langle 0\vert J(0)\vert\Gamma\rangle.
\end{equation} 
All the particles in the state $\vert\Gamma\rangle$ are
on--shell. This constrains the total energy--momentum $p_{\Gamma}$ to be a
time--like vector: $p_{\Gamma}^2=t$ with $t\ge 0$. With these constrains
on $p_{\Gamma}$ we can insert the identity
\begin{equation}
\int_{0}^{\infty}dt\int d^{4}p\,\theta(p)\,\delta (p^2 -t)\,\delta^{(4)}
(p-p_{\Gamma})=1
\end{equation} 
inside the sum $\sum_{\Gamma}$ over the complete set of
states. Interchanging the order of sum over $\Gamma$ and integration over
$t$ and $p$, there appears naturally the definition of the spectral
function associated with the
$J$--operator
\begin{equation}
\sum_{\Gamma}\langle 0\vert J(0)\vert\Gamma\rangle\langle\Gamma\vert
J(0)^{\dagger}\vert 0\rangle\,(2\pi)^4 \delta^{(4)}(p-p_{\Gamma})
\equiv 2\pi\rho(p^2).
\end{equation} 
The spectral function $\rho(p^2)$ is a scalar function of
the Lorentz invariant $p^2$ and the masses of the particles in the states
$\vert\Gamma\rangle$ only. By construction it is a real function and
non--negative
\begin{equation}
\rho(p^2)^* =\rho(p^2)\ge 0.
\end{equation}   
We can now rewrite the two--point function in
eq.~(\ref{eq:tpf}) as follows
\begin{equation}
\begin{split}
\Pi&(q^2)=\int d^4 x\,e^{iq\cdot x}\int_{0}^{\infty}dt\,\rho(t)\times\\
&\int
\frac{d^4 p}{(2\pi)^3}
\left[i\theta(x)\,e^{-ip\cdot
x}\theta(p)\delta(p^2-t)+i\theta(-x)\,e^{ip\cdot x}\theta(p)
\delta(p^2-t)\right].
\end{split}
\end{equation} 
Here, you may recognize the familiar functions of free field
theory
\begin{equation}
\Delta^{+}(x)=\int \frac{d^4 p}{(2\pi)^3} \,e^{-ip\cdot x}\theta(p)
\delta(p^2-t)
\end{equation} 
and
\begin{equation}
\begin{split}
\Delta^{-}(x) &= \int \frac{d^4 p}{(2\pi)^3}\,e^{ip\cdot
x}\theta(p)\delta(p^2-t) \\
&= \int \frac{d^4 p}{(2\pi)^3}\,e^{-ip\cdot x}\theta(-p)
\delta(p^2-t);
\end{split}
\end{equation} 
and therefore the Feynman propagator function
\begin{equation}
\begin{split}
\Delta_{\mbox{\small\rm F}}(x;t) &=
i\theta(x)\Delta^{+}(x;t)+i\theta(-x)\Delta^{-}(x;t)
\\
&=\int\frac{d^4 p}{(2\pi)^4}\frac{e^{-ip\cdot x}} {t-p^2-i\epsilon},
\end{split}
\end{equation} 
where the last expression can be obtained using the
representation in eq.~(\ref{eq:theta}) of the
$\theta$--function [{\footnotesize See e.g.
ref.~\cite{Veltman} pp. 50-51.}]. The two--point function $\Pi(q^2)$ appears
then to be the Fourier transform of a scalar free field propagating  with
an arbitrary mass squared $t$ weighted by the spectral function density
$\rho(t)$ and integrated over all possible values of $t$,
\begin{equation}
\Pi(q^2)=\int d^4 x\, e^{iq\cdot x}\int_{0}^{\infty}dt\,\rho(t)\,
\Delta_{\mbox{\small\rm F}}(x;t).
\end{equation} 
Integrating over $x$ and $p$ results finally in the wanted
representation
\begin{equation}
\Pi(q^2)=\int_{0}^{\infty}dt\,\rho(t)\frac{1}{t-q^2-i\epsilon}.
\end{equation} 
With 
\begin{equation}
\Pi(q^2)=\Ree\Pi(q^2)+i\Imm\Pi(q^2),
\end{equation} 
and the use of the identity in eq.~(\ref{eq:ppartid}), it
follows that
\begin{equation}
\rho(t)\equiv\frac{1}{\pi}\Imm\Pi(t),
\end{equation} 
which identifies the spectral function with the imaginary
part of the two--point function.

Notice that the formal manipulations above avoid the question of
convergence of the principal value integral
\begin{equation}
\Ree\Pi(q^2)=\mbox{\rm
PP}\!\int_{0}^{\infty}dt\,\frac{1}{t-q^2}\,\frac{1}{\pi}\Imm\Pi(t).
\end{equation} 
The convergence of the integral in the ultraviolet limit
($t\ra\infty$) depends on the behaviour of the spectral function at large
$t$--values. When doing above the exchange of sum over $\Gamma$ and
integrations we have implicitly assumed good convergence properties; but in
general the product of the distributions
$\theta(x)$ and
$\int_{0}^{\infty}dt\,\rho(t)\,\Delta^{+}(x;t)$ may not be a well--defined
distribution. The ambiguity manifests by the presence of an arbitrary
polynomial in $q^2$ in the r.h.s. of the PP--integral
\begin{equation}\label{eq:PPab}
\Ree\Pi(q^2)=\mbox{\rm PP}\!\int_{0}^{\infty}dt\,\frac{1}{t-q^2}\,
\frac{1}{\pi}\Imm\Pi(t) +\mbox{\rm a}+\mbox{\rm b}q^2 +\dots\, .
\end{equation} 
Notice that the coefficients of the arbitrary polynomial
have no discontinuities; in other words, the ambiguity of short--distance
behaviour reflects only in the evaluation of the real part of the
two--point function, not in the imaginary part. The physical meaning of
these coefficients depends of course on the choice of the local operator
$J(x)$ in the two--point function. In some cases the coefficients in
question are fixed by low--energy theorems; e.g. if
$\Pi(0)$ is known, we can trade the constant a in eq.~(\ref{eq:PPab}) for
$\Pi(0)$:
\begin{equation}
\Ree\Pi(q^2)=\Ree\Pi(0)+\mbox{\rm
PP}\!\int_{0}^{\infty}\frac{dt}{t}\frac{q^2}{t-q^2}\,
\frac{1}{\pi}\Imm\Pi(t) + \mbox{\rm b}q^2 + \dots\,.
\end{equation} 
In other cases the constants are absorbed by renormalization
constants. In general, it is always possible to get rid of the polynomial
terms by taking an appropriate number of derivatives with respect to
$q^2$. Various examples will be discussed later.

\subsection{QCD Moment Sum Rules}\label{subsec:msrs}

In QCD the number of derivatives which is required to obtain a well defined
two--point function is fixed by the asymptotic freedom property of the
theory. For a gauge invariant local operator
$J(x)$, the asymptotic behaviour of the associated two--point function is
of the type
\begin{equation}\label{eq:asymsp}
\lim_{t\ra\infty}\frac{1}{\pi}\Imm\Pi(t)\Ra {\mbox{\rm
A}}t^{d}\left\{1+a_{1}\frac{\als(t)}{\pi}+\dots\,\right\},
\end{equation} 
with A and $a_{1}$ calculable coefficients, and $d$ a known
integer
$d=0,1,2,\dots\,$, depending on the dimensions of the operator $J(x)$. It is
then sufficient to take
$d+1$ derivatives with respect to $q^2$ to get rid of the arbitrary
polynomial and obtain a convergent integral. The functions defined by the
{\it moment integrals} ($Q^2\equiv -q^2$):
\begin{equation}
\begin{split}
\Pi^{(m)}(Q^2) &= \frac{(-1)^m}{(m-d-1)!}\,(Q^2)^{m-d}\,
\frac{\partial^m}{(\partial Q^2)^m}\,\Pi(q^2) \\
&=\int_{0}^{\infty}dt\,\frac{m(m-1)\cdots
(m-d)}{(t+Q^2)^{d+1}}\,\left(\frac{Q^2}{t+Q^2}\right)^{m-d}
\frac{1}{\pi}\Imm\Pi(t),
\end{split}
\end{equation} 
for $m\ge d+1$ are then well defined functions calculable in
pQCD at sufficiently  large $Q^2$--values.

\subsection{The Adler Function}\label{subsec:adlerfunction}

An example of a QCD moment sum rule is the Adler function corresponding
to the two--point function associated with the vector--isovector quark
current
\begin{equation}
V^{\mu}_{\mbox{\scriptsize\rm I}=1}(x)=\frac{1}{2}\left[:\!{\bar
u}(x)\gamma^{\mu}u(x)\!:- :\!{\bar d}(x)\gamma^{\mu}d(x)\!:\right].
\end{equation}
Since the current is conserved, the two--point function
\begin{equation}\label{eq:adlsef}
\begin{split}
\Pi_{\mbox{\scriptsize\rm I}=1}^{\mu\nu}(q) &= i\int d^{4}x
e^{iq\cdot x}
\langle 0\vert\mbox{\rm T}
\left\{V_{\mbox{\scriptsize\rm I}=1}^{\mu}(x) V_{\mbox{\scriptsize\rm
I}=1}^{\nu}(0)\right\}\vert 0\rangle \\
 &= -(g^{\mu\nu}q^2 -q^{\mu}q^{\nu})\Pi_{\mbox{\scriptsize\rm
I}=1}(q^2),
\end{split}
\end{equation}
has one invariant function $\Pi_{\mbox{\scriptsize\rm I}=1}(q^2)$
only. Power counting tells us that in this case the asymptotic
behaviour of the corresponding spectral function goes as a constant
i.e. the power $d$ in the r.h.s. of eq.~(\ref{eq:asymsp}) is $d=0$.
Only one derivative is required to have a well defined correlation
function. The Adler function is defined as the logarithmic derivative
of
$\Pi(q^2)_{\mbox{\scriptsize\rm I}=1}$ ($Q^2\equiv -q^2$, with $Q^2\ge
0$ for
$q^2$--spacelike),
\begin{equation}
\cA (Q^2)\equiv -Q^2\frac{\partial \Pi_{\mbox{\scriptsize\rm
I}=1}(q^2)}{\partial Q^2}.
\end{equation}
This function, for sufficiently large values of $Q^2$ is calculable in
pQCD as a power series in the running coupling constant. On the
other hand the vector--isovector spectral function
$\frac{1}{\pi}\Imm\Pi_{\mbox{\scriptsize\rm I}=1}(t)$ is accessible to
experimental determination from
$e^{+}e^{-}$ annihilation into hadrons and from the $\tau$ hadronic
decays. We then have the sum rule
\begin{equation}
\cA(Q^2)=\int_{0}^{\infty}dt\,\frac{Q^2}{(t+Q^2)^2}\,
\frac{1}{\pi}\Imm\Pi_{\mbox{\scriptsize\rm I}=1}(t).
\end{equation}

When trying to confront this sum rule with experiment, there
appears the problem that the integrand in the r.h.s. is only
known experimentally from threshold up to finite values
of $t$. This brings in a question of {\it matching} whatever is
known about the low energy hadronic spectral function with its
asymptotic behaviour as predicted by pQCD. We shall later on come
back to this important question, after the discussion of
non--perturbative power corrections.

The Adler function  has been calculated in perturbation theory to the 
four-loop level~\cite{ChKT79,DS79,CG80,GKL91,SS91}. In the
$\overline{MS}$--scheme and  with
$\als\equiv \alpha_{\overline{MS}}({Q^2})$, the numerical
evaluation of the exact perturbative calculations for
$N_c =3$ and for three light flavours $n_f =3$ gives the result
\begin{equation}\label{eq:adlernumb}  
{\cal A} (Q^2)\vert_{\mbox{\scriptsize pert.}}  = 
\frac{N_c}{16\pi^2}\frac{2}{3}\left\{1+\frac{\alpha_{s}}{\pi}+
1.64\left(\frac{\alpha_{s}}{\pi}\right)^2+ 
6.37\left(\frac{\alpha_{s}}{\pi}\right)^3 \right\}.
\end{equation} 
       
\section{Types of Two--Point Function Sum Rules}\label{sec:ttpfsr}

There are several types of sum rules which have been discussed in the
literature. They differ on the way that the analyticity properties of
two--point functions are exploited and/or on various limits that one
can define on moment sum rules. We explain the most common types of sum
rules below.

\subsection{Laplace Transform Sum Rules}\label{subsec:latrsr}

The Laplace transform of a spectral function:
\begin{equation}\label{eq:laplace}
\cM(\sigma)=\int_{0}^{\infty}dt\,
e^{-t\sigma}\,\frac{1}{\pi}\Imm\Pi(t),
\end{equation} 
is a calculable function in pQCD provided that
the variable $\sigma$, which has the dimensions of an inverse mass
squared, is sufficiently small. The interesting point about this type of
sum rules is the presence of the exponential factor in
the integrand which gives a predominant weight to the low--energy
component of the hadronic spectral function compared to the
power weight of a moment sum rule. Sum rules of this type were first
proposed by the ITEP group~\cite{SVZ79} who called them Borel sum
rules. The Laplace transform in (\ref{eq:laplace}) results from a
precise limit of the moment sum rules discussed in
section~\ref{subsec:msrs}; it is the limit where the number of
derivatives $N$ of the invariant two--point function
$\Pi(q^2)$ with respect to $Q^2$ and $Q^2\equiv-q^2$  both go to infinity
with their ratio
$\frac{Q^2}{N}\equiv M^2\equiv\frac{1}{\sigma}$ fixed. More
precisely, the limit is obtained by applying the operator
\begin{equation}\label{eq:invlapop}
 {\mbox{\bf L}}\equiv\lim_{N\ra\infty}\lim_{Q^2\ra\infty}
\vert_{\frac{N}{Q^2}=\sigma}\,\frac{(-1)^{N}}{(N-1)!}
\left(Q^2\right)^{N}\frac{\partial^{N}}{\left(\partial Q^2\right)^{N}}
\end{equation}
to the first well--defined moment of $\Pi(q^2)$, [ $d$ is the
asymptotic power behaviour defined in eq.~(\ref{eq:asymsp})]:
\begin{equation}
\begin{split}
\frac{1}{\sigma} &{\mbox{\bf L}}\left[(-1)^{d+1}
\frac{\partial^{d+1}}{\left(\partial
Q^2\right)^{d+1}}\Pi(q^2)\right]=\\
&\int_{0}^{\infty}dt\,\frac{N(N+1)\cdots (N+d+1)}{(t+Q^2)^{d+1}}\,
\frac{1}{\left(1+\frac{1}{N}t\frac{N}{Q^2}\right)^{N}}\,
\frac{1}{\pi}\Imm\Pi(t).
\end{split}
\end{equation}
In the limit which we are considering, the first fraction in the
integrand goes to $\sigma^{d+1}$ and the second fraction to
$e^{-t\sigma}$, with the result that
\begin{equation}
\frac{1}{\sigma}{\mbox{\bf L}}\left[(-1)^{d+1}
\frac{\partial^{d+1}}{\left(\partial
Q^2\right)^{d+1}}\Pi(q^2)\right]
\Ra
\sigma^{d+1}\int_{0}^{\infty}dt\,e^{-t\sigma}
\,\frac{1}{\pi}\Imm\Pi(t).
\end{equation}

In practice, the question which then appears is how to calculate the
function $\cM (\sigma)$ without having to take the complicated limits
which define the operator ${\mbox{\bf L}}$ in
eq.~(\ref{eq:invlapop}). This is possible because
${\mbox{\bf L}}$ is nothing but an algebraic form
of the Laplace inversion operator. Once this is
recognized~\cite{BLR85}, it is easy to go directly from the moment sum
rule
\begin{equation}\label{eq:mom}
(-1)^{d+1}
\frac{\partial^{d+1}}{\left(\partial Q^2\right)^{d+1}}\Pi(q^2)=
\int_{0}^{\infty}dt\,\frac{(d+1)!}{(t+Q^2)^{d+2}}
\,\frac{1}{\pi}\Imm\Pi(t)
\end{equation}
to the Laplace transform in (\ref{eq:laplace}). Indeed, the power
weight in the integrand of eq.~(\ref{eq:mom}) can be written as a
Laplace transform with respect to the variable $Q^2$
\begin{equation}
\frac{(d+1)!}{(t+Q^2)^{d+2}}=\int_{0}^{\infty}
dx\,e^{-xQ^2}\,x^{d+1}\,e^{-xt},
\end{equation}
and the action of ${\mbox{\bf L}}$ brings back the inverse
Laplace transform i.e.,
\begin{equation}
{\mbox{\bf L}}\left[\frac{((d+1)!}{(t+Q^2)^{d+2}}\right]=
\sigma^{d+1}\,e^{-t\sigma}.
\end{equation}
This provides the key to calculate the Laplace transform
$\cM(\sigma)$ once we know $\Pi(q^2)$. The results of analytic pQCD
calculations of self--energy functions like $\Pi(q^2)$ are
combinations of terms which are typically of the form:
\begin{equation}
\left(\frac{\als(\nu^2)}{\pi}\right)^{p}\frac{1}{(Q^2)^{\alpha+1}}
\left\{1,\quad\log\frac{Q^2}{\nu^2},\quad
\log^2\frac{Q^2}{\nu^2},\quad\dots\right\}.
\end{equation}
The trick to go from $\Pi(q^2)$ to $\cM(\sigma)$ is to write
each of these terms as a Laplace transform with respect to the
variable $Q^2$. Since the operator
${\mbox{\bf L}}$ acts in fact as an inverse Laplace transform
one then finds e.g. that
\begin{equation}  
{\mbox{\bf
L}}\left[\frac{1}{(Q^2)^{\alpha+1}}\log\frac{Q^2}{\nu^2}\right]=
\frac{\sigma^{\alpha+1}}{\Gamma(\alpha+1)}\left(-\log\sigma\nu^2 
+\psi(\alpha+1)\right),
\end{equation}
where $\psi(z)=\frac{d}{dz}\log\Gamma(z)$.~[{\footnotesize Other
useful formulae like this one can be found e.g. in
ref.~\cite{BLR85}.}]

\begin{figure}[htb]
\centerline{\epsfbox{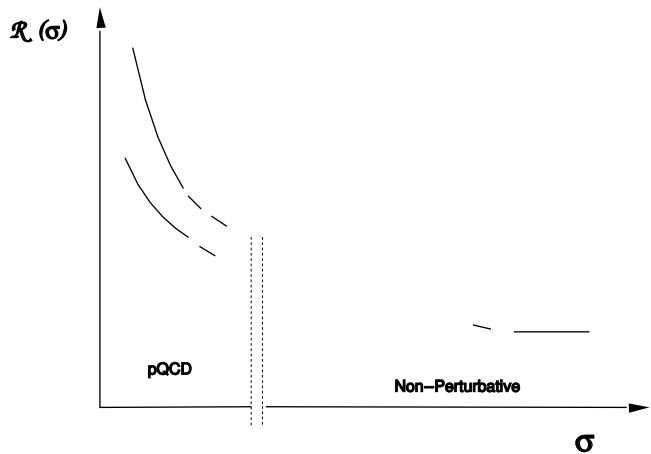}}
\caption{Expected shape of the short and long distance asymptotic behaviour 
of the function $\cR (\sigma)$.}
\label{fig:Fig3a}
\end{figure}

There are a number of interesting properties of the Laplace
transform which are useful to
know when doing phenomenological applications and therefore I
want to discuss them next. In sum rule applications one
often considers the ratio~\cite{BB81}
\begin{equation}
\cR(\sigma)=-\frac{d}{d\sigma}\log\cM(\sigma)=
\frac{\int_{0}^{\infty}dt\,e^{-t\sigma}\,t\,\frac{1}{\pi}\Imm\Pi(t)}
{\int_{0}^{\infty}dt\,e^{-t\sigma}\,\frac{1}{\pi}\Imm\Pi(t)}.
\end{equation}
At very small values of $\sigma$, the function $\cR (\sigma)$ has the
free--field behaviour
\begin{equation}
\cR (\sigma)\Ra\frac{d+1}{\sigma}\left(1+\cdots\right),
\end{equation}
where the dots indicate perturbative gluonic  corrections. Recall that
$d$ is the asymptotic power behaviour in $t$ of the corresponding spectral
function as shown in eq.~(\ref{eq:asymsp}). (For example $d=0$ in the case
where $J(x)$ is the vector--isovector current.) This dependence in $d$ is
the only reminiscence left from the subtractions needed in the dispersion
relation for the initial two--point function. At very large--$\sigma$ the
behaviour of
$\cR (\sigma)$ will be dominated by its non--perturbative behaviour.
We can get a feeling for this behaviour if we make a simple {\it
ansatz} for the hadronic spectral function, like e.g. the one
predicted in the large--$N_c$ limit of QCD where the hadronic spectral
function becomes an infinite sum of narrow states.~[{\footnotesize See
e.g. Manohar's lectures in this school~\cite{ManlesH}}.]
\begin{equation}
\frac{1}{\pi}\Imm\Pi(t)=\sum_{i}f_{i}^2 M_{i}^2\delta(t-M_{i}^2).
\end{equation}
It is easy to see that at large values of $\sigma$ the function
$\cR (\sigma)$ goes to the  mass squared of the ground state with
the quantum numbers of the current $J(x)$
\begin{equation}
\cR (\sigma)=\frac{\sum_{i}f_{i}^2 M_{i}^4\,e^{-M_{i}^2\sigma}}
{\sum_{i}f_{i}^2 M_{i}^2\,e^{-M_{i}^2\sigma}}\Ra M_{1}^2.
\end{equation}
These considerations give us a feeling of what is the expected
shape of the
$\cR$--function at asymptotic values of $\sigma$ . It corresponds to
the shape illustrated in Fig.~\ref{fig:Fig3a}.

\noi
The problem of course
is how to link the very short--distance regime (small $\sigma$)
to the very long--distance (large $\sigma$) one. We shall be able to
say a little bit more about that after the discussion of non--perturbative
power corrections in the next section.

I wish to conclude this subsection with the discussion of an
interesting {\it model independent} property of the
$\cR$--function~\cite{deR83}. It is the fact that, because of the
positivity property of a spectral function
$\frac{1}{\pi}\Imm\Pi(t)\ge 0$, the function $-\log\cM (\sigma)$ is a
concave function of
$\sigma$; or in other words, the slope of the function $\cR
(\sigma)$ must always be negative. This of course implies
severe restrictions on the way that the two asymptotic regimes
illustrated in Fig.~\ref{fig:Fig3a} can be joined. The proof of
this property is rather straightforward. It can be understood
very simply by making an analogy with statistical mechanics: $\cR
(\sigma)$ can be viewed as the equilibrium ``energy'' $\langle
t\rangle$ of a system with variable ``energy'' $t$ in thermal
equilibrium with a second system at ``temperature'' $1/\sigma$. In
this analogy, $\frac{1}{\pi}\Imm\Pi(t)$ represents the ``density of
states'' with ``energy'' $t$. Then the mean squared ``energy
fluctuation'' is given by
\begin{equation}
-\frac{d}{d\sigma}\cR (\sigma)\equiv \langle (t-\langle
t\rangle )^2\rangle = \langle t^2\rangle-\langle t\rangle^2 \ge
0,
\end{equation}
which by definition is a positive quantity.

\subsection{Gaussian Transform Sum Rules}

The Gaussian transform of a spectral function
\begin{equation}\label{eq:gausstr}
\cG(s,\tau)=\frac{1}{\sqrt{4\pi\tau}}\int_{0}^{\infty}dt\exp\left(
-\frac{(s-t)^2}{4\tau}\right)\frac{1}{\pi}\Imm\Pi(t),
\end{equation}  
can be evaluated in pQCD provided one keeps a finite
width $\sqrt{2\tau}$ resolution in the gaussian kernel  sufficiently large.
These sum rules provide the framework to formulate quantitatively the
notion of {\it local duality}. In the limit $\tau\ra 0$ where the gaussian
kernel becomes a delta function
\begin{equation}
\lim_{\tau\ra 0}\frac{1}{\sqrt{4\pi\tau}}
\exp\left( -\frac{(s-t)^2}{4\tau}\right)\Ra \delta (s-t),
\end{equation} 
the Gaussian transform coincides with the spectral function
itself
\begin{equation}
\lim_{\tau\ra 0} \cG(s,\tau)\Ra \frac{1}{\pi}\Imm\Pi(s).
\end{equation}  There is in fact an interesting analogy between these sum
rules and the theory of the ``heat equation''~\cite{BLR85}. The analogy is
based on the observation that $\cG(s,\tau)$ obeys the partial differential
equation
\begin{equation}
\frac{\partial^2 \cG(s,\tau)}{(\partial s)^2}=\frac{\partial
\cG(s,\tau)}{\partial \tau},
\end{equation} which is the one dimensional heat equation if we interpret
$s$ as a pseudo--``position'' variable and $\tau$ as a pseudo--``time''
variable.  In this analogy, the hadronic spectral function corresponds to
the initial ``heat'' (or ``temperature'') distribution in a semi--infinite
rod
$0\le s\le\infty$ and
$\cG(s,\tau)$ measures the evolution in ``time'' $\tau$ of the ``heat'' 
distribution in this rod. The calculation of $\cG(s,\tau)$ in pQCD is an
expansion in powers of $\alpha_{s}(\sqrt{\tau})$ which blows up at small
$\tau$--values; therefore, direct confrontation of the physical hadronic
spectral function $\frac{1}{\pi}\Imm\Pi(t)$ with pQCD cannot be made at
small $\tau$--values.   The confrontation can however be made at
sufficiently large
$\tau$--values by comparing the ``evolved'' physical hadronic spectral
function
$\cG(s,\tau)$ with the corresponding pQCD theoretical prediction.  The more
we knew about non--perturbative corrections the smaller we should be able to
go in ``time'' i.e. the smaller we would be able to take the width
$\sqrt{2\tau}$ of the gaussian kernel and therefore the closer one could
satisfy {\it local duality}.

\begin{figure}[htb]
\centerline{\epsfbox{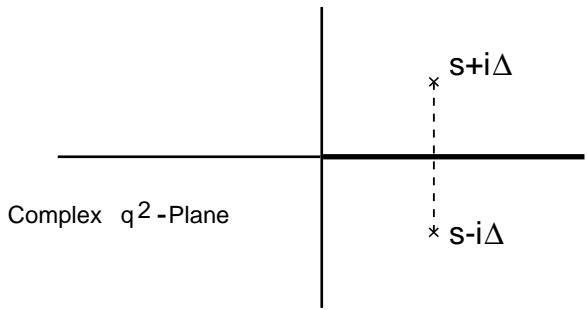}}
\caption{Points in the complex $q^2$--plane where the two--point function in
eq.~(\ref{eq:lorentz}) is evaluated.}
\label{fig:Fig3b}
\end{figure}

Let us discuss how to get the Gaussian transform from a generic two--point
function like $\Pi(q^2)$ in eq.~(\ref{eq:tpf}). First, one evaluates
$\Pi(q^2)$ at a complex point $q^2=s+i\Delta$ ($s$ and $\Delta$ are real
positive variables) and at its complex conjugate $q^2=s-i\Delta$ (see
Fig.~\ref{fig:Fig3b})

\noi 
and defines the combination, (I am assuming for
simplicity that the dispersion relation for $\Pi(q^2)$ requires at most one
subtraction, but the argument can be easily generalized as in the case
discussed for the Laplace transform.)
\begin{equation}\label{eq:lorentz}
\frac{\Pi(s+i\Delta)}{i\Delta}+\frac{\Pi(s-i\Delta)}{-i\Delta}=
\int_{0}^{\infty} dt\,\frac{1}{(t-s)^2 +\Delta^2}\,\frac{1}{\pi}\Imm (t).
\end{equation} The integral in the r.h.s. brings in the convolution with a
Lorentz--like kernel which we can write as a Laplace transform
\begin{equation}
\frac{1}{(t-s)^2 +\Delta^2}=\int_{0}^{\infty}dx\,e^{-x\Delta^2}\,
e^{-x(t-s)^2}.
\end{equation} 
Applying to this integral representation the techniques developed in the
previous section~\ref{subsec:latrsr} allows us to construct the inverse
Laplace transform operator which is needed to obtain the Gaussian transform
in eq.~(\ref{eq:gausstr}) from the Lorentz transform in
eq.~(\ref{eq:lorentz}). It is the operator
\begin{equation}
{\mbox{\bf L}}\equiv\lim_{N\ra\infty}\,\lim_{\Delta^2\ra\infty}
\left\vert_{\frac{1}{N}\Delta^2=4\tau}\right.\,\frac{(-1)^{N}}{(N-1)!}
\left(\Delta^2\right)^{N}\frac{\partial^{N}}{\left(\partial
\Delta^2\right)^{N}}.
\end{equation} 
We then have the desired relation
\begin{equation}
\begin{split}
\frac{1}{\sqrt{4\pi\tau}} &2\tau {\mbox{\bf L}} 
\left[\frac{\Pi(s+i\Delta)}{i\Delta}+\frac{\Pi(s-i\Delta)}{-i\Delta}
\right]\\
&\Ra
\frac{1}{\sqrt{4\pi\tau}}\int_{0}^{\infty}dt\exp\left(
-\frac{(s-t)^2}{4\tau}\right)\frac{1}{\pi}\Imm\Pi(t).
\end{split}
\end{equation}
I will later come back to the physics of Gaussian sum rules, but I need
first to introduce some properties of finite energy sum rules which is the
subject of the next section.

\subsection{Finite Energy Sum Rules}\label{subsec:sfesr}

\begin{figure}[htb]
\centerline{\epsfbox{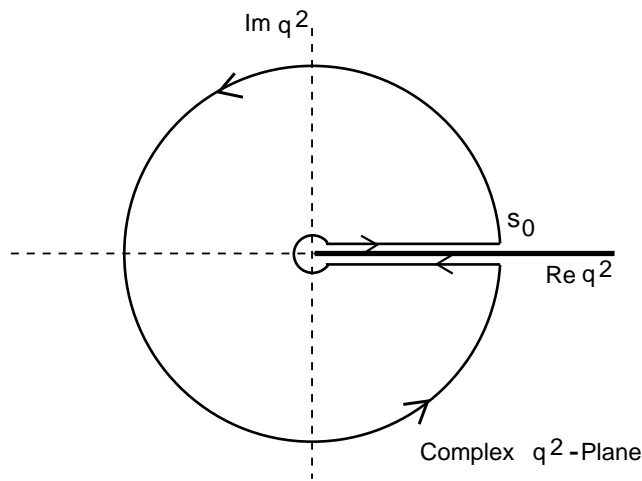}}
\caption{Contour of integration in the complex $q^2$--plane.}
\label{fig:Fig3c}
\end{figure}

The moments of spectral functions
\begin{equation}
\cM_{n}(s)=\int_{0}^{s_{0}}dt\,t^{n}\,\frac{1}{\pi}\Imm\Pi(t),
\end{equation} with $n=0,1,2,\cdots$ are also calculable quantities in QCD,
provided that the upper limit $s_{0}$ is sufficiently large. Finite energy
sum rules follow from the analyticity properties of two--point functions
already discussed in section~\ref{sec:disperrels}. Applying Cauchy's
integral formula to the contour in the complex $q^2$--plane shown in
Fig.~\ref{fig:Fig3c}

\noi
 avoids the cut in the real axis and therefore results 
in
\begin{equation}
\frac{1}{2\pi i}\oint dq^2\,(q^2)^{n}\,\Pi(q^2)=0.
\end{equation} We now separate this integral into two pieces: one is the
contribution from the paths above and below the real axis which pick up the
discontinuity of the
$\Pi(q^2)$ function along this axis, and hence its imaginary part; the
other is the contribution to the integral over the circle of radius
$s_{0}$~[{\footnotesize One should also check that there is no contribution
from the little circle around the origin. The behaviour of $\Pi(q^2)$ at
small $q^2$ is usually known from low--energy properties of the two--point
function.}]. These two contributions have to match, resulting in the equation
\begin{equation}\label{eq:fesrsn}
\int_{0}^{s_{0}}dt\,t^n\, \Imm\Pi(t)=-\frac{1}{2\pi i}
\oint_{\vert q^2 \vert=s_{0}}dq^2\,q^{2n}\,\Pi(q^2).
\end{equation} 
The integral in the l.h.s. is obtained using hadronic data,
or inserting an hadronic {\it ansatz}; the integral in the r.h.s. is done
in pQCD integrating the running coupling constant dependence over the
circle of radius $s_{0}$ and scaling the result at $\mu^2 =s_{0}$.
Non--perturbative
$1/q^2$--power corrections can also contribute in general to the r.h.s.
integral. In QCD, and with $d$ the asymptotic power behaviour of a given
spectral function as defined in eq.~(\ref{eq:asymsp}), finite energy sum
rules with $n+d\ge 1$ are particularly sensitive to non--perturbative power
corrections. These corrections will be the subject of the next section. 

In finite energy sum rules, the question of subtractions in the primitive
two--point functions can be dealt with by an appropriate integration by
parts. For example, if one subtraction in $\Pi(q^2)$ is required as for
example in the case of the Adler function discussed in
section~\ref{subsec:adlerfunction}, and we want to compute a general moment
integral over the the circle of radius
$s_{0}$, we can use the identity ($f(q^2)$ is assumed to be non--singular,)
\begin{equation}
\oint_{\vert q^2 \vert=s_{0}} dq^2\,f(q^2)\,\Pi(q^2)= -\oint_{\vert q^2
\vert=s_{0}}\frac{dq^2}{q^2}\left[F(q^2)-F(s_{0})\right]
q^2\frac{d\Pi(q^2)}{dq^2},
\end{equation} 
with $F(z)=\int_{0}^{z}dw\,f(w)$.  As one increases the
$n$--power in a finite energy sum rule, one becomes more and more sensitive
to the detailed high energy behaviour of the hadronic spectral function. In
the QCD expression of the r.h.s., this is reflected by the appearence of
higher and higher values of the coefficients in the perturbative series in
$\alpha_{s}(s)$ and therefore worse convergence, as well as by the appearance of
non--perturbative condensates of higher and higher dimension.

Finite energy sum rules provide a very useful way to do the correct
matching between a given {\it ansatz} of the low--energy hadronic spectral
function and the onset of the perturbative continuum. For example, in the
case of the Adler function discussed in section~\ref{subsec:adlerfunction},
let us consider the simplest hadronic {\it ansatz} of a narrow low--energy
ground state plus a continuum i.e. the {\it ansatz} in
eq.~(\ref{eq:Vresdomqcd}) which we have already discussed in connection
with the phenomenological study of the Weinberg sum rules:
\begin{equation}\label{eq:vmdansatz}
\frac{1}{\pi}\Imm\Pi_{\mbox{\scriptsize\rm I}=1}(t)=f_{\rho}^2
M_{\rho}^2\delta(t-M_{\rho}^2)+
\frac{N_c}{16\pi^2}\frac{2}{3}\theta(t-s_{0})[1+\cdots].
\end{equation}     
In this case, the application of the finite energy sum
rule in eq.~(\ref{eq:fesrsn}) with $n=0$ and in the chiral limit where $m_u
=m_d\ra 0$ does not depend on non--perturbative condensates (because of the
absence of gauge invariant operators of dimension $d=2$ in QCD,) and results
in the following constraint among the parameters of the hadronic {\it
ansatz}
\begin{equation}\label{eq:vmdfesr}  
f_{\rho}^2
M_{\rho}^2=\frac{N_c}{24\pi^2}\,s_{0}[1+\cdots].
\end{equation} 
Numerically, using the experimental values of
$f_{\rho}$ and
$M_{\rho}$ in this equation results in a value for the onset of the
perturbative continuum
$s_{0}\simeq 1.8\,\GeV^2$ which is quite reasonable, in the sense that it is
sufficiently large to trust pQCD~[{\footnotesize With gluonic corrections
incorporated at the two loop level, the resulting $s_{0}$ is reduced to
$s_{0}\simeq 1.6\,\GeV^2$}.]. 
        
\begin{figure}[htb]
\centerline{\epsfbox{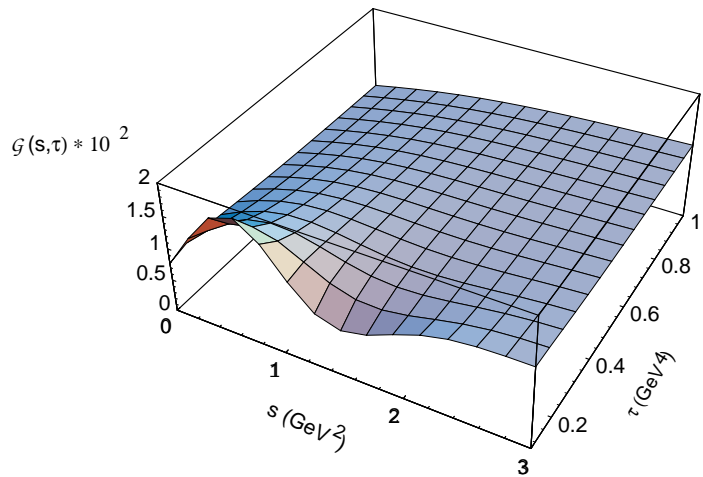}}
\caption{The Gaussian Transform of the Spectral Function in
eq.~(\ref{eq:vmdansatz}) .}
\label{fig:Fig3d}
\end{figure}

\begin{figure}[htb]
\centerline{\epsfbox{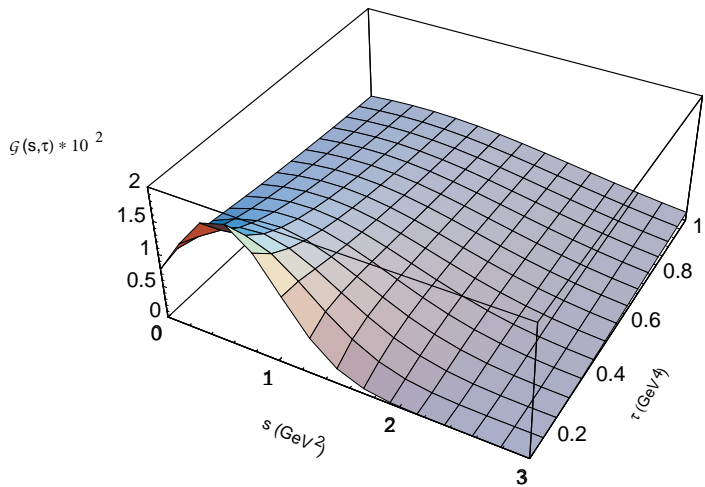}}
\caption{The Gaussian Transform of the Spectral Function in
eq.~(\ref{eq:specdelta}) .}
\label{fig:Fig3e}
\end{figure}

I can now comment on a very interesting connection between finite energy
sum rules and the Gaussian sum rules we discussed in  the previous
subsection. It has been shown in a general way~\cite{BLR85}, that matching
a given low--energy hadronic spectral function {\it ansatz} to the
perturbative continuum via finite energy sum rules guarantees that the
corresponding Gaussian transform at large widths has then the correct
asymptotic behaviour predicted by pQCD. This is an important point in
phenomenological applications and therefore I want to discuss it in some
detail. Very often one finds in the literature rather ``precise'' QCD sum
rule ``predictions'' of masses or couplings which, however, have been
obtained by fixing the onset of the pQCD continuum in the hadronic spectral
function at some value using some ``good physical sense'' criteria, or even
forgetting altogether about the QCD continuum. You can check that,
unfortunately, the Gaussian transform of these spectral functions do not
evolve to the shape predicted by pQCD. In other words, parameterizations of
spectral functions which are not correctly constrained by finite energy sum
rules are not guaranteed to be {\it globally dual} to QCD and therefore
should be avoided. Let us illustrate this with some pictures.
Figure~\ref{fig:Fig3d} shows the evolution in the pseudo--``time'' variable
$\tau$ of the Gaussian transform of the spectral function {\it ansatz} in
eq.~(\ref{eq:vmdansatz}) with the onset of the continuum $s_{0}$ fixed by
the finite energy sum rule in eq.~(\ref{eq:vmdfesr}). 
\noi In the ``heat evolution'' analogy the spectral
function in eq.~(\ref{eq:vmdansatz}) corresponds to the initial ``heat
distribution'' in the $s$--axis. The picture shows the evolution in
``time'' of this ``heat distribution'' in the interval $0.1\,\GeV^4 \le\tau
\le 1\,\GeV^4$.  We
observe that asymptotically in ``time'', i.e. for 
$\tau$ large, the spectral function evolves very well to the asymptotic
``heat distribution'' predicted by pQCD i.e.
\begin{equation}
\lim_{\tau\ra\infty}\cG(s,\tau)\Ra\frac{1}{16\pi^2}
\left(1-\mbox{\rm erf}(\frac{s}{2\sqrt{\tau}})\right)[1+\cdots],
\end{equation}  where $\mbox{\rm erf}(x)$ denotes the error function
$\mbox{\rm erf}(x)=\frac{2}{\sqrt{\pi}}\int_{0}^{x} dy\,e^{-y^2}$. By
contrast, Fig.~\ref{fig:Fig3e} shows the same evolution in the limit case
of only a delta--function {\it ansatz} for the spectral function
\begin{equation}\label{eq:specdelta}
\frac{1}{\pi}\Imm\Pi_{\mbox{\scriptsize\rm I}=1}(t)=f_{\rho}^2
M_{\rho}^2\delta(t-M_{\rho}^2),
\end{equation}
 with no continuum.
\noi 
Clearly the corresponding asymptotic
``heat distribution'' fails to reproduce the shape predicted by pQCD.
{\it Global duality} of a given hadronic spectral function {\it ansatz} with
QCD is only obtained provided that the hadronic parameters are
constrained to satisfy a system of finite energy sum rules equations.

\section{Non--Perturbative Power Corrections}\label{sec:nppc}

The physical vacuum of QCD is not the vacuum state which one uses in
perturbation theory. Physical effects like spontaneous chiral symmetry
breaking and/or confinement do not appear in an order by order perturbative
treatment of QCD. A natural question to ask then is {\it how are the pQCD
results modified by non--perturbative effects at short--distances?} This is
the question which we shall be dealing with in this section. We shall see
that non--perturbative effects in two--point functions evaluated at large
$Q^2$--values appear as inverse power corrections in $Q^2$. They are the
analogue of the {\it higher twist } corrections of deep inelastic
scattering~\cite{ManlesH}. They can be systematically evaluated by doing
Wilson's  {\it operator product expansion}~\cite{Wil69} (OPE) in the
physical vacuum.  The power corrections appear then as the product of Wilson
coefficients, which are calculable perturbatively, times universal
non--zero vacuum expectation values of gauge invariant operators, the so
called {\it condensates} which, although excluded by definition order by
order in perturbation theory, they can {\it a priori} have non--zero values
in the physical vacuum. Typical examples are the lowest dimension {\it
quark condensate} $\langle\bar{\psi}\psi\rangle$ and {\it gluon condensate}
$\gs^2\langle\tr G_{\mu\nu}G^{\mu\nu}\rangle$. The operators which appear
in any of the vacuum condensates which we shall be considering are assumed
to be {\it normal ordered}.

\subsection{Infrared Renormalon Ambiguities and Power Corrections}

\begin{figure}[htb]
\centerline{\epsfbox{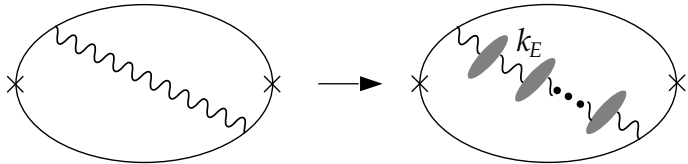}}
\caption{One of the Feynman diagrams which contributes to the Adler
function with the gluon propagator replaced by an effective charge chain.}
\label{fig:Fig4a}
\end{figure}

\begin{figure}[htb]
\centerline{\epsfbox{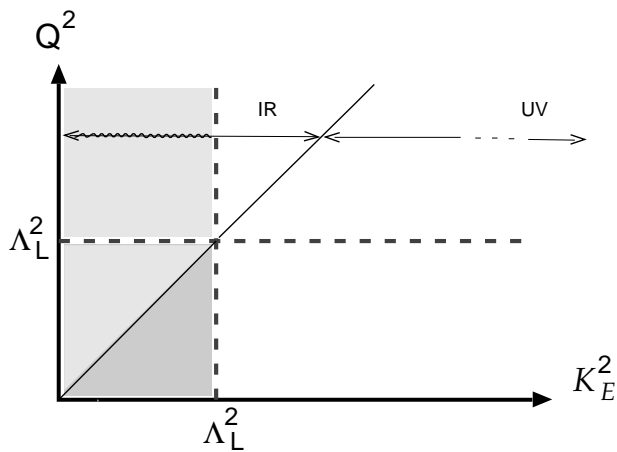}}
\caption{Integration regions of the virtual euclidean momentum $k_{E}^2$
for a fixed euclidean external momenta
$Q^2$.}
\label{fig:Fig4b}
\end{figure}

The need of non--perturbative power corrections to colour singlet gauge
invariant two--point functions can be ``hinted'' from what emerges already
in perturbation theory when summing a specific class of contributions. Let
us consider the calculation of a typical two--point function, e.g. the
Adler function which we have already defined in
section~\ref{subsec:adlerfunction}, and let us restrict our attention to
the class of Feynman diagrams generated by the replacement of a virtual
gluon by one {\it effective charge} chain as shown in Fig~\ref{fig:Fig4a}.
\noi Formally, the chain of bubbles in Fig.~\ref{fig:Fig4a} corresponds to
the replacement of the ordinary free gluon propagator by the  full {\it
gluon propagator--like} function 
\begin{equation}\label{eq:gr}
\begin{split}
&\frac{-i\left(g_{\mu\nu}-(1-\xi)
\frac{k_{\mu}k_{\nu}}{k^2}\right)}{k^2 +i\epsilon}(-ig_{s})^2
\Rightarrow \\
&-i\left(g_{\mu\nu}-\frac{k_{\mu}k_{\nu}}{k^2}\right)
\frac{4\pi\,\alpha_{\mbox{\small{\rm eff}}}(k_E^2)}{k_{E}^2-i\epsilon}
 + i\xi\frac{k_{\mu}k_{\nu}}{k^2}
\frac{(-ig_s)^2}{k^2 +i\epsilon},
\end{split}
\end{equation} 
where $k_{E}$ denotes the euclidean virtual momentum carried
by the chain;  and $\xi$ is the usual covariant gauge parameter. The
interesting quantity here is the {\it effective charge} function
$\alpha_{\mbox{\small{\rm eff}}}(k_{E}^2)$. The existence of an {\it
effective charge} in non--abelian gauge theories with properties precisely
analogous to those of the well known {\it effective charge} of QED is
discussed in refs.\cite{Wat97,PRW97} and references therein. In terms of a
characteristic scale, like e.g. the
$\Lambda_{\overline{MS}}$ scale of the
$\overline{MS}$--renormalization scheme,
\begin{equation}\label{eq:master}
\alpha_{\mbox{\small{\rm eff}}}(k_{E}^2)=\frac{1}{\frac{-\beta_{1}}{2\pi}
\log\frac{k_E^2}{c^2\Lambda_{\overline{MS}}^2}},
\end{equation} where
\begin{equation}
\beta_{1}=\frac{-11}{6}N_c +\frac{1}{3}n_f\qquad
\mbox{\rm and} \qquad c^2=\exp\left\{\frac{67N_c -10n_f}{33N_c
-6n_f}\right\}\,.
\end{equation} The $\beta_{1}$ factor is precisely the first coefficient of
the QCD
$\beta$--function. The {\it effective charge} encodes therefore the physics
of the $\beta$--function; in particular the scale breaking property. Since,
eventually, we are interested in the appearance of physical scales in QCD,
it seems natural to focus our attention on the properties of those diagrams
of perturbation theory generated by the insertion of {\it effective
charge}--exchanges. Of course, we do that at the simplest level of keeping
only the one loop dependence of the $\beta$--function. As far as one is
only  interested in general qualitative features this should not be a
serious limitation. It is helpful now to look at the
$Q^2$--$k_{E}^2$ plot in Fig.~\ref{fig:Fig4b}, where $k_{E}$ denotes the 
euclidean energy--momentum carried by the chain in Fig.~\ref{fig:Fig4a}.
\noi  We shall refer to the integration regions $0\le k_{E}^{2}\le Q^2$ and 
$Q^2 \le k_{E}^2\le\infty$ shown in Fig.~\ref{fig:Fig4b} as the IR--region
and the UV--region respectively. It is this IR--region of integration which
is at the origin of the appearance of singularities, the so called IR
renormalons, when tree level gluon propagators are replaced by the full
summation of perturbation theory diagrams which define a {\it renormalon
chain} as illustrated in Fig.~\ref{fig:Fig4a}. For reference, we also show
in the plot in Fig.~\ref{fig:Fig4b} the lines
$Q^2=\Lambda_{L}^2$ and
$k_{E}^2=\Lambda_{L}^2$, with $\Lambda_L=c\times
\Lambda_{\overline{MS}}\simeq 1\,\GeV$ the {\it Landau scale} at which the
pQCD {\it effective charge} becomes singular. In pQCD calculations, the
external euclidean momenta is always chosen to be
$Q^2\gg\Lambda_{L}^2$; however, regardless of how large
$Q^{2}$ is taken, there will always be a region in the virtual
$k_{E}^2$ integration which is below $\Lambda_{L}^2$ where pQCD is not
defined. It is precisely this region of integration which, as we shall see
in the next paragraph, leads to IR renormalon poles in the so called Borel
plane. The appearance of these singularities is welcome because they
reflect the limitations of perturbation theory and indicate the presence of
non--perturbative power corrections, as indeed the OPE in the physical 
vacuum suggests~[{\footnotesize The reader may be curious to know about the
fate of the UV renormalons in QCD. Ref.~\cite{PdeR97} and the references
therein discuss this interesting topic.}].

The leading contribution to the self--energy function
$\Pi_{\mbox{\scriptsize\rm I}=1}(q^2)$ in eq.~(\ref{eq:adlsef}) from the
chain exchange diagrams in Fig.\ref{fig:Fig4a} in the IR domain of
integration $0\le k_{E}^{2}\le Q^2$ can be easily obtained and it has the
following form $\left(\CF=\frac{N_c^2 -1}{2N_c}\right)$
\begin{equation}
\Pi_{\mbox{\scriptsize\rm I}=1}(Q^2)\Ra
\frac{N_c}{32\pi^2}\CF\int_{0}^{Q^2}dk_{E}^{2}\frac{k_{E}^{2}}{Q^2}
\alpha_{\mbox{\small{\rm eff}}}(k_{E}^2).
\end{equation} Making the change of variables
\begin{equation}  w/2=- b_0
\alphaq \log k_E^2/Q^2\,,\qquad \mbox{\rm where}\qquad b_0\equiv
-\beta_1/2\pi  
\end{equation}  in the previous equation, one obtains 
\begin{equation}
\Pi_{\mbox{\scriptsize\rm I}=1}(Q^2)\vert_{\mbox{\small{\rm IR}}}
=\frac{N_c}{16\pi^2}\,\CF\,\frac{1}{2\pi b_0}
\int_{0}^{\infty}dw\, e^{-\frac{w}{b_0\alpha(\mu^2)}}\:\frac{1}{2-w}
\left(\frac{Q^2}{\mu^2}\right)^{-w},
\end{equation} which leads to a contribution to the Adler function
\begin{equation}\label{eq:irr}
\cA(Q^2)\vert_{\mbox{\small{\rm IR}}}
=\frac{N_c}{16\pi^2}\,C_{F}\,\frac{1}{2\pi b_0}
\int_{0}^{\infty}dw\, e^{-\frac{w}{b_0\alpham}}\:\frac{w}{2-w}
\left(\frac{Q^2}{\mu^2}\right)^{-w}.
\end{equation}
\noi  This expression is already in the form of a Borel transform. An
expansion around $w=0$ would generate the characteristic $n!$ behaviour of
the perturbative expansion in $\alpham$ indicating that the corresponding
series is not Borel summable. As already discussed by other
authors~\cite{Mu85,Za92,Be93}, we find that the leading IR renormalon
contribution to the Adler function appears as a pole in the Borel plane  at
the location $w=2$. There is no term in the IR expansion of the domain of
integration which leads to a pole at
$w=1$. From the previous change of variables one also sees that low values 
of
$w$ are associated with momenta around the large scale $Q^2$, where
perturbation theory is expected to give a good description of the  dynamics.
However, as $w$ goes up (and it has to go all the way up to infinity) one 
enters deeper and deeper into the IR region, where perturbation theory must 
fail or else QCD would not describe the spectrum of hadronic bound states
that are  observed. The singularity at $w=2$ exhibits this in its crudest
form.  More  precisely, this singularity implies an ambiguity in the 
perturbative  evaluation  of the Adler function. Therefore, the analytic
continuation in $w$ in eq.~(\ref{eq:irr}), or equivalently the resummation
of perturbation theory into the effective charge $\alphak$ of
eq.~(\ref{eq:master}), that  one has tried in order to obtain the full
solution has failed. The ambiguity  is encoded in the unavoidable 
prescription to skip the pole; and is of the form $\sim e^{-1/b_0
\alpha(Q)}$ :  
\begin{equation}\label{eq:ambiguity}
\Delta\Pi(Q^2)\vert_{\mbox{\small{\rm
IR}}}\sim\frac{N_c}{16\pi^2}\,\frac{6}{11}\,
\frac{N_{c}^2 -1}{N_{c}^2-2n_{f}}
\left(\frac{\Lambda_L}{Q}\right)^4,
\end{equation}
\noi i.e., an ambiguity which has the same $\frac{1}{Q^4}$ pattern as the
gluon condensate contribution  which appears in the OPE evaluation of  the
Adler function~\cite{Da82,Da84,NSVZ84,NSVZ85}. Since the Adler function
must be an unambiguous physical observable, there must exist another
contribution that cancels the one in eq.~(\ref{eq:ambiguity}).  In other
words, (Borel) resummed perturbation theory is requiring the presence of
the gluon condensate  (which is also ambiguous for the same reason) to
combine with the result of eq.~(\ref{eq:irr}) and yield a final
well--defined answer~\cite{Mu85}.   This is an example of how all--orders
perturbation theory is capable of ``hinting'' at  nonperturbative dynamics.

\subsection{The ITEP Power Corrections}

\begin{figure}[htb]
\centerline{\epsfbox{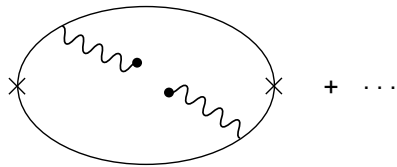}}
\caption{Feynman diagram representation of gluon
condensate contributions.}
\label{fig:Fig4c}
\end{figure}

The intuitive physical picture which lies behind the non--perturbative
power corrections introduced by the ITEP group~\cite{SVZ79} can be
understood as follows. In terms of Feynman diagrams, the configurations
which are sensitive to the IR-region are those where the hard external
momentum
$Q$ flows into the diagram and goes along the quark loop with very little
transfer to the gluon propagator, or the one--renormalon chain. The gluon
propagator is then in a non--perturbative regime where many soft gluon
interactions are possible since the coupling constant in this regime is
very large. The idea of the OPE is to factor out the hard part of the
$Q$--flow and calculate it perturbatively, while collecting the bulk of the
soft non--perturbative gluon interactions in a  {\it condensate} of gluons
which will be treated as a phenomenological parameter.  In practice, the
calculations in the OPE are performed by integrating the fermion loop in
the presence of ``background'' gluon fields, including hard perturbative
gluonic corrections if necessary~[{\footnotesize  Ref.\cite{Noetal84} gives
a lot of technical details on how to do these calculations.}]. The average of
all possible gauge invariant configurations of ``background'' gluon fields
defines then the {\it condensates}. Background fields in an ordinary
Feynman diagram are indicated by terminating dots, like in
Fig.~\ref{fig:Fig4c}.
\noi The dots are a short--hand notation for the non--perturbative average
over soft interactions. The perturbative evaluation of the ``hard'' part of
the loop gives the corresponding Wilson coefficient and the net result is a
contribution of the type
$\cC_{GG}\frac{\gs^2\langle G_{a}^{\mu\nu}G^{a}_{\mu\nu}\rangle}{Q^4}$,
 where $\cC_{GG}$ denotes the Wilson coefficient. The Wilson coefficient
depends on the specific two--point function one is calculating, while the
{\it gluon condensate}
$\gs^2\langle G_{a}^{\mu\nu}G^{a}_{\mu\nu}\rangle$ is treated as a universal
phenomenological parameter. For example, in the case of the Adler function
discussed in section~\ref{subsec:adlerfunction} the {\it gluon condensate}
gives a correction to the perturbative result in eq.~(\ref{eq:adlernumb}):
\begin{equation}
\cA (Q^2)\vert_{GG}=\frac{1}{12}\frac{\als}{\pi}\frac{\langle
G_{a}^{\mu\nu}G^{a}_{\mu\nu}\rangle}{Q^4}.
\end{equation} 

\begin{figure}[htb]
\centerline{\epsfbox{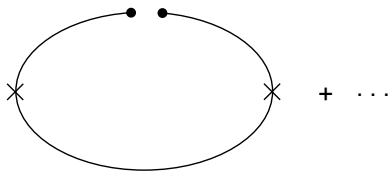}}
\caption{Feynman diagram representation of quark condensate contributions.}
\label{fig:Fig4d}
\end{figure}

For two--point functions with composite operators of light quarks, there
are also $1/Q^4$--corrections induced by the {\it quark condensate}, as
illustrated in Fig.~\ref{fig:Fig4d} below
\noi  which give contributions of the type
$\cC_{{\bar\psi}\psi}\frac{m\langle{\bar\psi}_{L}\psi_{R}+
{\bar\psi}_{R}\psi_{L}\rangle}{Q^4}$, where, as before, the Wilson
coefficient
$\cC_{{\bar\psi}\psi}$ depends on the specific current operator. The
appearance of a mass factor in front of the {\it quark condensate} is due
to the conservation of chirality. In the case of the Adler function
discussed in section~\ref{subsec:adlerfunction} the {\it quark condensate}
gives a correction to the perturbative result in eq.~(\ref{eq:adlernumb}):
\begin{equation}
\cA (Q^2)\vert_{{\bar\psi}\psi}=\frac{m_u \langle{\bar u}u\rangle+ m_d
\langle{\bar d}d\rangle}{Q^4}.
\end{equation}

\begin{figure}[htb]
\centerline{\epsfbox{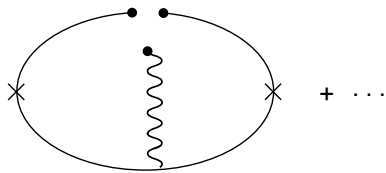}}
\caption{Feynman diagram representation of contributions from the mixed
quark gluon condensate.}
\label{fig:Fig4e}
\end{figure}

\begin{figure}[htb]
\centerline{\epsfbox{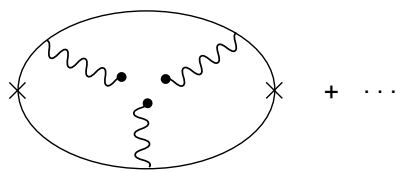}}
\caption{Feynman diagram representation of contributions from the three
gluon condensate.}
\label{fig:Fig4f}
\end{figure}

\begin{figure}[htb]
\centerline{\epsfbox{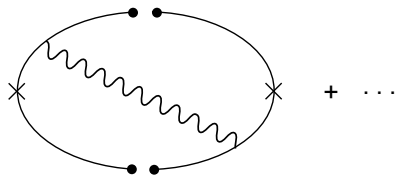}}
\caption{Feynman diagram representation of contributions from the four
quark condensate.}
\label{fig:Fig4g}
\end{figure}

Higher order terms in the OPE bring in contributions from condensates of
higher and higher dimension. Three new types of condensates appear at $\cO
(1/Q^6)$: the {\it quark gluon mixed condensate} illustrated in
Fig.~\ref{fig:Fig4e} below
\noi which brings contributions of the type
$\cC_{{\bar\psi}\psi}\frac{m
\gs\langle{\bar\psi}_{L}\sigma^{\mu\nu}\lambda^{(a)}\psi_{R}
G_{\mu\nu}^{(a)}\rangle}{Q^6}$; the {\it three gluon condensate}
illustrated in Fig.~\ref{fig:Fig4f}
\noi which can give contributions like
$\cC_{GGG}\frac{\langle\gs^3
G_{\mu\nu}^{(a)}G^{(b)\nu}_{\;\rho}G^{(b)\rho\mu}\rangle}{Q^6}$; and the
{\it four quark condensate} illustrated in Fig.~\ref{fig:Fig4g}
\noi which contributes to terms like
$\cC_{{\bar\psi}\psi{\bar\psi}\psi}\frac{\langle{\bar\psi}\Gamma\psi
{\bar\psi}\Gamma\psi\rangle}{Q^6}$. In the case of the Adler function
discussed in section~\ref{subsec:adlerfunction} the {\it quark gluon mixed
condensate} is suppressed by $\frac{m^2}{Q^2}$ factors. It has been
shown~\cite{DuSm81,HuMa82} that the {\it three gluon condensate}, without
extra hard gluon corrections, gives no contribution to the Adler function.
The {\it four quark condensate} contribution was first calculated in
ref.\cite{SVZ79} and it  gives a correction to the perturbative result in
eq.~(\ref{eq:adlernumb}):
\begin{equation}
\begin{split}
\cA
(Q^2)_{{\bar\psi}\psi{\bar\psi}\psi} &=\left[-\frac{3}{2}\left\langle\left(
{\bar u}\gamma_{\mu}\gamma_{5}\lambda^{a}u- {\bar
d}\gamma_{\mu}\gamma_{5}\lambda^{a}d\right)^2\right\rangle\right.\\
&-\left.
\frac{1}{3}\left\langle\left( {\bar u}\gamma_{\mu}\lambda^{a}u+{\bar
d}\gamma_{\mu}\lambda^{a}d\right)\sum_{q=u,d,s}{\bar
q}\gamma^{\mu}\lambda^{a}q\right\rangle\right]
\frac{\pi\als}{Q^6}.
\end{split}
\end{equation} 
To leading order in the $1/N_c$--expansion these {\it four
quark condensates} factorize into products of the lowest dimension {\it
quark condensate}.

The phenomenological determinations of the {\it gluon condensate} and the
{\it quark condensate} have still rather big errors. A generous range for
the {\it gluon condensate} which covers most of the determinations is
\begin{equation}
\langle\als G_{a}^{\mu\nu}G^{a}_{\mu\nu}\rangle=(0.08\pm 0.04)\,\GeV^4.
\end{equation} The {\it quark condensate} is scale dependent. The values
obtained from a variety of sum rules are within the range
\begin{equation}
\frac{\langle\bar{u}u+\bar{d}d\rangle(1\,\GeV^2)}{2}=
-(0.013\pm0.004)\,\GeV^3.
\end{equation} 
It turns out that with those values for the lower dimension
{\it condensates}, the contribution from the leading non--perturbative power
corrections in most of the sum rule applications -at the
$Q^2$ values where one can trust the pQCD contributions-  turn out to be
already relatively small. This is due to the fact that the scale of
perturbation theory
$\Lambda_{\overline{MS}}$ is rather large; therefore, in order to get a
convergent series in $\als(Q^2)$ one is obliged to go to rather high values
of $Q^2$ where the power corrections are already quite small.

\subsection{The Weinberg Sum Rules in the Light of the
OPE}\label{subsec:wsrope}

We are now in the position to understand why it is that the Weinberg sum
rules we discussed in section~\ref{subsec:wsrss} are satisfied in QCD.
For this it is sufficient to consider the OPE applied to the two--point
function
$\Pi_{LR}(q)$ defined in eq.~(\ref{eq:lrtpf})~[{\footnotesize Notice that
the OPE of this self--energy function is free from renormalon ambiguities,
because in perturbation theory it is protected by chiral symmetry, which
ensures that $\Pi_{LR}(Q^2)=0$ order by order in powers of $\alpha_{s}$.}].
Since this two--point function is chirally symmetric when quark masses are
set to zero, the only possible inverse powers of $Q^2$ which can appear are
those modulated by local order parameters of S$\chi$SB. In QCD there are no
local order parameters of dimensions $d=2$ and $d=4$. Notice that
$\langle\bar{\psi}\psi\rangle$, which has dimensions $d=3$ contributes to
$\frac{1}{Q^4}$--terms but it appears multiplied by a quark mass factor; 
and therefore it disappears in the chiral limit. The first possible
contribution comes from local order parameters of dimension $d=6$, i.e. the
four--quark condensate contributions of the type illustrated in
Fig.~\ref{fig:Fig4f}. The specific contribution to $\Pi_{LR}(Q^2)$ has
been calculated in ref.~\cite{SVZ79}. Their result is particularly simple
in the large--$N_c$ limit where the vacuum expectation value of four--quark
operators factorizes in a product of
$\langle\bar{\psi}\psi\rangle$ terms, and therefore we have that
\begin{equation}\label{eq:OPELR}
\lim_{Q^2\ra\infty}\Pi_{LR}(Q^2)=\frac{1}{Q^6}
\left[-4\pi^2\left(\frac{\alpha_s}{\pi}+\cO(\alpha_s^2)\right)
\langle\bar{\psi}\psi\rangle^2
\right]+\cO\left(\frac{1}{Q^8}\right)\,.
\end{equation} From this follow the two constraints in eqs.~(\ref{eq:ope2})
and (\ref{eq:ope4}), and hence the two Weinberg sum rules in
eqs.~(\ref{eq:1stwsr}) and (\ref{eq:2ndwsr}).

We also observe that the $\Pi_{LR}(Q^2)$ expression in
eq.~(\ref{eq:lrtpfrd}) which results from the resonance dominance ansatz
for the vector and axial--vector spectral functions has the required
$\frac{1}{Q^6}$ behaviour at large--$Q^2$
\begin{equation}
\lim_{Q^2\ra\infty}\Pi_{LR}(Q^2)=\left(f_{V}^2 M_{V}^6 -f_{A}^2
M_{A}^6\right)\frac{1}{Q^6}+\cO(\frac{1}{Q^8}).
\end{equation} This is not surprising since the ansatz in question is
nothing but a specific choice  compatible with the spectrum predicted by
QCD in the large--$N_c$ limit.

In section~\ref{subsec:wsrss} I stressed the fact that it was not enough to
know the the very low $Q^2$ behaviour of $\Pi_{LR}(Q^2)$ in order to
calculate the integral in eq.~(\ref{eq:piem}) which gives the electromagnetic
$\pi^{+}-\pi^{0}$ mass difference. We now observe that it is not enough
either to know only the very high $Q^2$ behaviour of the same function. The
$\frac{1}{Q^6}$ behaviour of the short--distance OPE in
eq.~(\ref{eq:OPELR}) cannot be extrapolated to the infrared region of the
integral since it would lead to a divergent result as well. The simple
resonance dominance ansatz in eq.~(\ref{eq:lrtpfrd}) provides however, in
this case, a correct matching of short--distance behaviour with the
long--distance behaviour.  

\section{Some Examples of QCD Sum Rules}\label{sec:seqcdsr}

Here we shall discuss two applications of the  techniques about sum rules
in QCD developed in the previous sections. From a phenomenological point
of view, the most striking application is the determination of the QCD
coupling constant $\als(m_{\tau}^{2})$ which has been made using the LEP
data on hadronic tau decays. Another recent application uses very general
properties of QCD sum rules to set lower bounds on the light quark masses. 

\subsection{Determination of the QCD Coupling Constant from Tau
decays}

Hadronic tau decays are governed by the spectral function associated with the
charged current 
$L^{\mu}(x)=\bar{u}(x)\gamma^{\mu}\frac{1}{2}(1-\gamma_{5})d_{\theta}(x)$,
where $d_{\theta}$ denotes the Cabibbo rotated $d$--quark. From the QCD
point of view, the relevant two--point function is then
\begin{equation}\label{eq:LLtpf}
\begin{split}
\Pi_{LL}^{\mu\nu}(q) & =4i\int d^4 x\,e^{iq\cdot x}\langle 0\vert
\mbox{\rm T}\left(L^{\mu}(x)L^{\nu}(0)^{\dagger}\right)\vert 0\rangle
\\
 & =(q^{\mu}q^{\nu}-g^{\mu\nu}q^2)\Pi_{LL}^{(1)}(Q^2)+
q^{\mu}q^{\nu}\Pi_{LL}^{(0)}(Q^2),
\end{split}
\end{equation}
which in full generality, when the quark masses are not neglected has
two invariant functions corresponding to physical states with total angular
momentum $J=1$ and $J=0$. From the phenomenological point of view, the
quantity which is accessible to experimental determination is the hadronic
tau decay branching ratio 
\begin{equation}\label{eq:rtauexp}
\begin{split}   
R_{\tau} &\equiv
\frac{\Gamma(\tau\ra\nu_{\tau}+\mbox{\rm hadrons})}
{\Gamma(\tau^{-}\ra\nu_{\tau}e^-
\bar{\nu}_{e})}=12\pi\times \\ 
&\int_{0}^{m_{\tau}^2}\frac{dt}
{m_{\tau}^2}\left(1-\frac{t}{m_{\tau}^2}\right)^2 
\left\{\left(1+2\frac{t}{m_{\tau}^2}
\right)\Imm\Pi_{LL}^{(1)}(t)+\Imm\Pi_{LL}^{(0)}(t)\right\}.
\end{split}
\end{equation}
The basic idea to apply here is the one we explained in the
section~\ref{subsec:sfesr} on finite energy sum rules, i.e. to confront the
experimental determination of $R_{\tau}$ with the corresponding theoretical
result~\cite{BNP92} obtained from the contour integral in the complex
$q^2$--plane shown in Fig.~\ref{fig:Fig3c} with $s_{0}=m_{\tau}^2$:
\begin{equation}\label{eq:rtauth}
\begin{split} 
R_{\tau}=6\pi i\times \oint_{\vert
q^2\vert=m_{\tau}^2} &\frac{dq^2}
{m_{\tau}^2}\left(1-\frac{q^2}{m_{\tau}^2}\right)^2 \\ 
&\times\left\{\left(1+2\frac{q^2}{m_{\tau}^2}
\right)\Pi_{LL}^{(1)}(q^2)+\Pi_{LL}^{(0)}(q^2)\right\}.
\end{split}
\end{equation}  
The analyticity properties of the two--point function in
eq.~(\ref{eq:LLtpf}) imply that the r.h.s.'s of eqs.~(\ref{eq:rtauexp}) and
(\ref{eq:rtauth}) should be equal. Notice that the evaluation of this
integral only requires the knowledge of the invariant functions
$\Pi_{LL}^{(1)}(q^2)$ and $\Pi_{LL}^{(0)}(q^2)$ in the complex big circle of
radius
$\vert q^2\vert=m_{\tau}^2$ and
$m_{\tau}\simeq 1.78\,\GeV$ is a reasonably large scale to apply pQCD.
Non--perturbative power corrections to this integral bring inverse powers
of $m_{\tau}^2$. In fact, one of the nice features about this integral is
that because of the combination of $q^2$--powers in the phase space factor
which modulate the
$\Pi_{LL}(q^2)$'s functions in the integrand, the leading $1/m_{\tau}^4$
non--perturbative power corrections are suppressed up to terms of
$\cO\left[\left(\frac{\als(m_{\tau}^2)}{\pi}\right)^2\frac{\langle\als
GG\rangle}{m_{\tau}^4}\right]$. Also the contribution from the region in
the circle near the real axis, where the validity of the OPE is perhaps
questionable, is suppressed by the kinematic factor
$\left(1-\frac{q^2}{m_{\tau}^2}\right)^2$.  The result of the explicit
calculation, which invokes quite  a lot of refined techniques~\cite{LeP92},
can be written as follows
\begin{equation}  
R_{\tau}=N_c\left(\vert V_{ud}\vert^2 +\vert
V_{us}\vert^2\right)^2 S_{\mbox{\rm\footnotesize
EW}}\left\{1+\delta^{'}_{\mbox{\rm\footnotesize EW}}+\delta^{(0)}+
\delta_{\mbox{\rm\footnotesize NP}}\right\},
\end{equation} 
where $S_{\mbox{\rm\footnotesize EW}}=1.0194$ and
$\delta^{'}_{\mbox{\rm\footnotesize EW}}=0.0010$ are the contributions from
the leading and next--to--leading electroweak corrections, and
\begin{equation}
\begin{split}
\delta^{(0)}= &\frac{\als(m_{\tau}^2)}{\pi}+
5.2023\left(\frac{\als(m_{\tau}^2)}{\pi}\right)^2
+26.366\left(\frac{\als(m_{\tau}^2)}{\pi}\right)^3 \\
&+\cO\left(\frac{\als(m_{\tau}^2)}{\pi}\right)^4,
\end{split}
\end{equation}  
is the result of the pQCD calculation in the chiral limit.
The remaining factor $\delta_{\mbox{\rm\footnotesize NP}}\approx -0.016\pm
0.005$ includes the estimated effects of small quark mass--corrections and
non--perturbative power corrections.

The most recent analysis, using the ALEPH data~\cite{Hoc97} gives the value
\begin{equation}
\als(m_{\tau})=0.348\pm0.008_{\mbox{\scriptsize exp}}\pm
0.015_{\mbox{\scriptsize th}}.
\end{equation}  
This value, when evolved with the renormalization group
running to the
$Z$--mass scale gives
\begin{equation}
\als(M_Z)=0.1211\pm 0.0008_{\mbox{\scriptsize exp}}
\pm 0.0016_{\mbox{\scriptsize th}}\pm 0.0010_{\mbox{\scriptsize evol}},
\end{equation} 
which, with these errors, represents the most accurate
determination of
$\als(M_Z)$ from a single experiment .

\subsection{Lower Bounds on the Light Quark Masses}

The light  $u$,  $d$ and $s$ quarks in the QCD Lagrangian have bilinear
couplings
\begin{equation}\label{lagrangian}
\cL_{\QCD}=\cL_{\QCD}^{(0)}-\sum_{q=u,d,s}m_{q}{\bar q}(x)q(x)\,,
\end{equation} 
with masses $m_q\neq 0$ which explicitly break the chiral
symmetry of the Lagrangian. The present situation concerning the values of
the light quark masses in the $\overline{\mbox{\rm MS}}$ renormalization
scheme, or any other scheme for that matter, has become rather confusing 
because of recent determinations of the light quark masses from lattice
QCD~\cite{BG96,Mactal96} which find substantially lower values than those
previously  obtained using a variety of QCD sum rules, (see e.g.
refs.~\cite{BPR95,JM95,CDPS95} and earlier references therein.) These new
lattice determinations are also in disagreement with earlier lattice
results~\cite{Alletal94}. More recently, an independent lattice QCD
determination~\cite{Eietal97} using dynamical Wilson fermions finds results
which, for $m_{u}+m_{d}$ are in agreement within errors with those of
refs.~\cite{BG96,Mactal96}; while for
$m_{s}$ they agree rather well, again within errors, with the QCD sum rule
determinations. [{\footnotesize For a very lucid introduction to the
techniques of Lattice QCD, see the lectures of M.~L\"{u}scher at this
school~\cite{Luch97}.}] Here, and as an application of the topics discussed
in the previous sections, I shall show that there exist rigorous lower
bounds on how small the light quark masses
$m_{s}+m_{u}$ and
$m_{d}+m_{u}$ can be~\cite{LRT97}. There are two--point functions
\begin{equation}\label{eq:2pf}
\Psi (q^2)\equiv i\int d^{4}x e^{iq\cdot x}\langle 0\vert T\left(\cO (x)\cO
(0)^{\dagger}\right)\vert 0\rangle,
\end{equation} 
with an appropriate choice of the local operator $\cO(x)$
which are particularly sensitive to quark masses: the case where $\cO$ is
the divergence of the strangeness changing axial current
\begin{equation}\label{eq:dac}
\cO (x)\equiv\partial_{\mu}A^{\mu}(x)=(m_{s}+m_{u}):\!\!{\bar
s}(x)i\gamma_{5}u(x)\!\!:;
\end{equation}   
and another one where $\cO$ is the scalar operator $S(x)$,
defined as the isosinglet component of the mass term in the QCD Lagrangian
\begin{equation}\label{eq:scc} 
\cO (x)\equiv S(x)=\hat{m}[:\!\!{\bar u}(x)u(x)\!\!:+:\!\!{\bar
d}(x)d(x)\!\!:],
\end{equation}  
with
\begin{equation}
\hat{m}\equiv\frac{1}{2}(m_u + m_d).
\end{equation} 
Both two--point functions obey dispersion relations which in
QCD require two subtractions, and it is therefore appropriate to consider
their second derivatives ($Q^2\equiv -q^2$):
\begin{equation}\label{eq:disp}
\Psi''(Q^2)=\int_{0}^{\infty} dt\frac{2}{(t+Q^2)^3}\frac{1}{\pi}\Imm\Psi(t).
\end{equation}  
The bounds follow from the restriction of the sum over all
possible hadronic states which can contribute to the spectral
function to the state(s) with the lowest invariant mass. It turns out that,
for the two operators in (\ref{eq:dac}) and (\ref{eq:scc}), these hadronic
contributions are well known phenomenologically; either from experiment or
from chiral perturbation theory ($\chi$PT) calculations. On the QCD side of
the dispersion relation, the two--point functions in question where the
quark masses appear as an overall factor, are known in the deep euclidean
region from pQCD at the four loop level. The leading non--perturbative
$\frac{1}{Q^2}$--power corrections which appear in the operator product
expansion when the time ordered product in (\ref{eq:2pf}) is evaluated in
the physical vacuum~\cite{SVZ79} are also known. For the sake of
simplicity, I shall only discuss here the case where $\cO$ is the
divergence of the strangeness changing axial current in eq.~(\ref{eq:dac}).

We shall call $\Psi_{5}(q^2)$ the two--point function in (\ref{eq:2pf})
associated with the divergence of the strangeness changing axial current
in (\ref{eq:dac}). The lowest hadronic state which contributes to the
corresponding spectral function is the $K$--pole. From eq.~(\ref{eq:disp})
we then have
\begin{equation}\label{eq:psi5}
\Psi_{5}''(Q^2)= \frac{2}{(M_{K}^2 +Q^2)^3}2f_{K}^2 M_{K}^4 +
\int_{t_{0}}^{\infty} dt
\frac{2}{t+Q^2)^3}\frac{1}{\pi}\mbox{Im}\Psi_{5}(t),
\end{equation}  
where $t_{0}=(M_{K}+2m_{\pi})^2$ is the threshold of the hadronic 
continuum.

It is convenient to introduce the moments $\Sigma_{N}(Q^2)$ of the hadronic
continuum integral
\begin{equation}\label{eq:contint}
\Sigma_{N}(Q^2)=\int_{t_{0}}^{\infty}dt\frac{2}{(t+Q^2)^3}\times 
\left(\frac{t_0 +Q^2}{t+Q^2}\right)^{N}\frac{1}{\pi}\Imm\Psi_{5}(t).
\end{equation}  
One is then confronted with a typical moment problem (see
e.g. ref.~\cite{AhK62}.) The positivity of the continuum spectral function
$\frac{1}{\pi}\Imm\Psi_{5}(t)$ constrains the moments
$\Sigma_{N}(Q^2)$ and hence the l.h.s. of (\ref{eq:psi5}) where the light
quark masses appear. The most general constraints among the first three
moments for
$N=0,1,2$ are:
\begin{equation}
\Sigma_{0}(Q^2)\ge 0,\quad \Sigma_{1}(Q^2)\ge 0,\quad
\Sigma_{2}(Q^2)\ge 0;
\end{equation}
\begin{equation}\label{eq:diff}
\Sigma_{0}(Q^2)-\Sigma_{1}(Q^2)\ge 0,\quad
\Sigma_{1}(Q^2)-\Sigma_{2}(Q^2)\ge 0;
\end{equation}
\begin{equation}
\label{eq:quad}
\Sigma_{0}(Q^2)\Sigma_{2}(Q^2)-\left(\Sigma_{1}(Q^2)\right)^2\ge 0\,.
\end{equation}   
The inequalities in eq.~(\ref{eq:diff}) are in fact trivial
unless
$2Q^2< t_{0}$, which constrains the region in $Q^2$ to too small values
for pQCD to be applicable. The other inequalities lead however to
interesting bounds which we next discuss.

The inequality $\Sigma_{0}(Q^2)\ge 0$ results in a first bound on the
running masses:
\begin{equation}\label{eq:1stbound}
\begin{split}
\left[m_{s}(Q^2)+m_{u}(Q^2)\right]^2 \ge &\frac{16\pi^2}{N_c}
\frac{2f_{K}^2 M_{K}^4}{Q^4} \\ &\times
\frac{1}{\left(1+\frac{M_{K}^2}{Q^2}\right)^3}
\frac{1}{\left[1+
\frac{11}{3}\frac{\alpha_{s}(Q^2)}{\pi} +\cdots\right]},
\end{split}
\end{equation} 
where the dots represent higher order terms which have been
calculated up to
${\cal O}(\alpha_s^3)$, as well as  non--perturbative power corrections of
$\cO\left(\frac{1}{Q^4}\right)$ and strange quark mass corrections of
$\cO\left(\frac{m_{s}^2}{Q^2}\right)$ and
$\cO\left(\frac{m_{s}^4}{Q^4}\right)$ including ${\cal O}(\alpha_s)$
terms~.   Notice that this bound is non--trivial in the large--$N_c$ limit
($f_{K}^2\sim\cO(N_c)$) and in the chiral limit ($m_{s}\sim M_{K}^2$). The
bound is of course a function of the choice of the euclidean
$Q$--value at which the r.h.s. in eq.~(\ref{eq:1stbound}) is evaluated. For
the bound to be meaningful, the choice of $Q$ has to be made sufficiently
large. In ref.~\cite{LRT97} it is shown that
$Q\gtrsim 1.4\:\GeV$ is already a safe choice to trust the pQCD
corrections as such. The lower bound which follows from
eq.~(\ref{eq:1stbound}) for
$m_u + m_s$ at a renormalization scale
$\mu^2=4\:\GeV^2$ results in the solid curves shown in Fig.~\ref{fig:Fig5a}
below. 
\begin{figure}[htb]
\centerline{\epsfbox{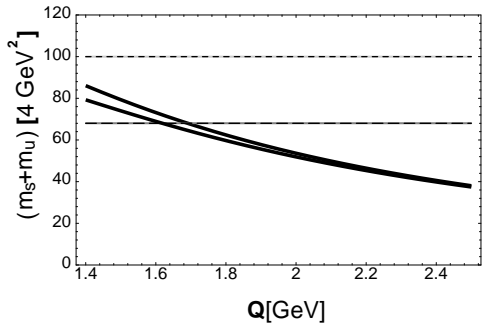}}
\caption{Lower bound in {\rm MeV} for
$[m_{s}+m_{u}](4\:\protect\GeV^2)$ versus $Q(\protect\GeV)$ from
eq.~(\protect\ref{eq:1stbound}).}
\label{fig:Fig5a}
\end{figure}
\noi  These are the lower bounds obtained by letting
$\Lambda^{(3)}_{\overline{MS}}$ vary~\cite{PDB} between 290~MeV  (the upper
curve) and 380~Mev (the lower curve). Values of the quark masses below the
solid curves in Fig.~\ref{fig:Fig5a} are forbidden by the bounds. For
comparison, the horizontal lines in the figure correspond to the central
values obtained by the authors of ref.~\cite{BG96}: their ``quenched''
result is the horizontal upper line; their ``unquenched'' result the
horizontal lower line. It must be emphasized that in Fig.~\ref{fig:Fig5a}
one is comparing what is meant to be a ``calculation'' of the quark masses
--the horizontal lines which are the lattice results of ref.~\cite{BG96}--
with a bound  which in fact can only be saturated in the limit where
$\Sigma_{0}(Q^2)= 0$. Physically, this limit corresponds to the extreme case
where the hadronic spectral function from the continuum of states is totally
neglected! What the plots show is that, even in that extreme limit, and  for
values of $Q$ in the range
$1.4\:\GeV\lesssim Q\lesssim 1.6\:\GeV$ the ``unquenched'' lattice results
of ref.~\cite{BG96} are already in difficulty with the bounds. 

The quadratic inequality in (\ref{eq:quad}) results in improved lower
bounds for the quark masses which we show in Fig.~\ref{fig:Fig5b} below. 
\begin{figure}[htb]
\centerline{\epsfbox{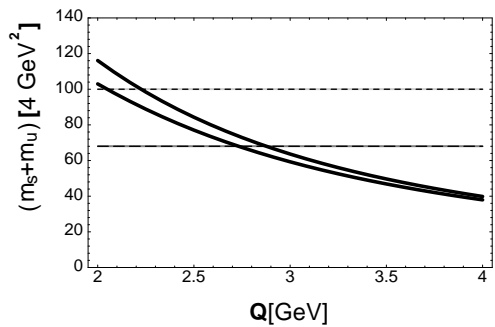}}
\caption{Lower bound in MeV for
$[m_{s}+m_{u}](4\:\protect\GeV^2)$ from the quadratic inequality.}
\label{fig:Fig5b}
\end{figure}
\noi Like in Fig.~\ref{fig:Fig5a}, these are the lower bounds obtained by
letting
$\Lambda^{(3)}_{\overline{MS}}$ vary~\cite{PDB} between 290~MeV (the upper
curve) and 380~Mev (the lower curve). Values of the quark masses below the
solid curves in Fig.~\ref{fig:Fig5b} are forbidden by the bounds. The
horizontal lines in this figure are also the same lattice results as in
Fig.~\ref{fig:Fig5a}.

The quadratic bound is saturated for a $\delta$--like spectral function
representation of the hadronic continuum of states at an arbitrary
position and with an arbitrary weight. This is certainly less restrictive
than the extreme limit with the full hadronic continuum neglected, and it is
therefore not surprising that the quadratic bound happens to be better than
the ones from
$\Sigma_{N}(Q^2)$ for
$N=0,1,$ and $2$. Notice however that the quadratic bound in
Fig.~\ref{fig:Fig5b} is plotted at higher
$Q$--values than the bound in Fig.~\ref{fig:Fig5a}. This is due to the fact
that the coefficients of the perturbative series in $\als(Q^2)$ become
larger for the higher moments. In ref~\cite{LRT97} it is shown that for the
evaluation of the  quadratic bound $Q\gtrsim 2\:\GeV$ is already a safe
choice. One finds  that in this case even the quenched lattice results of
refs.~\cite{BG96,Mactal96} are in difficulty with these bounds.

Similar bounds can be obtained for $m_{u}+m_{d}$ when one considers the
two--point function associated with the divergence of the axial current
\begin{equation}
\partial_{\mu}A^{\mu}(x)=(m_{d}+m_{u}):\!\!{\bar d}(x)i\gamma_{5}u(x)\!\!:.
\end{equation}   The method to derive the bounds is exactly the same as the
one discussed above and therefore we only show, in Fig.~\ref{fig:Fig5c}
below, the results for the corresponding lower bounds  which one obtains
from the quadratic inequality.
\begin{figure}[htb]
\centerline{\epsfbox{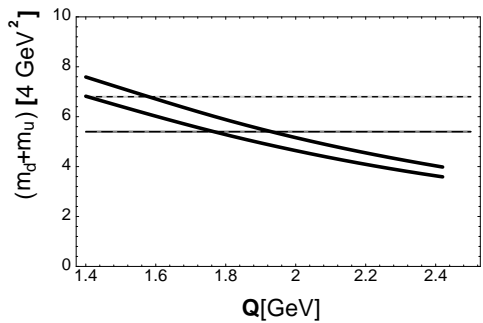}}
\caption{Lower bound in  MeV for
$[m_{d}+m_{u}](4\:\protect\GeV^2)$ from the quadratic inequality.}
\label{fig:Fig5c}
\end{figure}
\noi   One finds again that the lattice QCD results of
refs.~\cite{BG96,Mactal96,Eietal97} for $m_{u}+m_{d}$ are in serious
difficulties with these bounds.
The lower bounds  discussed above are perfectly compatible with the sum
rules results of refs.~\cite{BPR95,JM95,CDPS95} and earlier references
therein.

The two examples discussed in this section are rather illustrative  of how
sum rules can relate QCD short--distance calculations  to
low--energy hadronic experimental information. There are  many other
examples discussed in the literature. My hope is that these introductory
lectures may help you to read the literature and to find new applications
as well.   

\vspace*{0.5cm}
\noi {\bf Acknowledgements}

\noi While preparing these lectures I have benefited a lot from the advice
and help of my colleagues Santi Peris, Michel Perrottet and Joaquim Prades.
It is a pleasure to thank them here. I also want to thank the students at
the Les Houches summer school for their interest and active participation.
They have been a very positive stimulus in producing this write--up.

\end{document}